\documentclass[twocolumn,journal]{IEEEtran} %

\title{6G Radio Testbeds: Requirements, Trends, and Approaches}
\author{Gilles Callebaut, Liang Liu, Thomas Eriksson, Liesbet Van der Perre, Ove Edfors and Christian Fager}

\usepackage[utf8]{inputenc}
\usepackage[switch]{lineno} %
 \usepackage[acronym,shortcuts]{glossaries}
\makeglossaries
\newacronym{3gpp}{3GPP}{3rd Generation Partnership Project}
\newacronym{6g}{6G}{sixth-generation}
\newacronym{5g}{5G}{fifth-generation}
\newacronym{lo}{LO}{local oscillator}
\newacronym{mate}{MATE}{millimeter-wave MIMO testbed}
\newacronym{if}{IF}{intermediate-frequency}
\newacronym{sfp}{SFP}{small form-factor pluggable}
\newacronym{abp}{ABP}{Authentication By Personalisation}
\newacronym{aclr}{ACLR}{adjacent channel leakage ratio}
\newacronym{adr}{ADR}{Adaptive Data Rate}
\newacronym{ag}{AG}{array gain}
\newacronym{ai}{AI}{Artificial Intelligence}
\newacronym{amam}{AM/AM}{amplitude modulation to amplitude modulation}
\newacronym{amp}{AMP}{approximate message passing}
\newacronym{ampm}{AM/PM}{amplitude modulation to phase modulation}
\newacronym{aoa}{AOA}{angle-of-arrival}
\newacronym{aod}{AOD}{angle-of-departure}
\newacronym{ap}{AP}{access point} 
\newacronym{ar}{AR}{augmented reality}
\newacronym{vr}{VR}{virtual reality}
\newacronym{xr}{XR}{extended reality}
\newacronym{arp}{ARP}{Antenna Reference Point}
\newacronym{asic}{ASIC}{application specific integrated circuit}
\newacronym{auv}{AUV}{autonomous underwater vehicle}
\newacronym{awgn}{AWGN}{additive white Gaussian noise}
\newacronym{apu}{APU}{access point unit}
\newacronym{bom}{BOM}{bill of materials}
\newacronym{bs}{BS}{base station}
\newacronym{bldc}{BLDC}{brushless DC}
\newacronym{bw}{BW}{Bandwidth}
\newacronym{bf}{BF}{beamforming}
\newacronym{bb}{BB}{base-band}
\newacronym{baw}{BAW}{bulk acoustic wave}
\newacronym{cars}{CARS}{calibration reference signal}
\newacronym{cbm}{CBM}{Condition Based Maintenance}
\newacronym{cc}{CC}{constant current}
\newacronym{ccnn}{CCNN}{circular convolutional neural network}
\newacronym{cir}{CIR}{channel impulse response}
\newacronym{cdf}{CDF}{cumulative distribution function}
\newacronym{cdrx}{CDRX}{Connected Mode DRX}
\newacronym{ce}{CE}{coverage enhancement}
\newacronym{cost}{COST}{commercial off-the-shelf}
\newacronym{ced}{CED}{cumulative energy density}
\newacronym{cf}{CF}{cell-free}
\newacronym{cfo}{CFO}{carrier frequency offset}
\newacronym{cots}{COTS}{commercial off-the-shelf}
\newacronym{cp}{CP}{cyclic prefix}
\newacronym{cpt}{CPT}{capacitive power transfer}
\newacronym{cpu}{CPU}{central-processing unit}
\newacronym{cqi}{CQI}{channel quality indicator}
\newacronym{cr}{CR}{Coding Rate}
\newacronym{crlb}{CRLB}{Cram\'er-Rao lower bound}
\newacronym{crs}{CRS}{cell reference signal}
\newacronym{cs}{CS}{compressed sensing}
\newacronym{csi}{CSI}{channel state information}
\newacronym{csp}{CSP}{contact service point}
\newacronym{css}{CSS}{Chirp Spread Spectrum}
\newacronym{cv}{CV}{constant voltage}
\newacronym{dac}{DAC}{digital-to-analog converter}
\newacronym{adc}{ADC}{analog-to-digital converter}
\newacronym{dc}{DC}{direct current}
\newacronym{dcc}{DCC}{dynamic cooperation clustering}
\newacronym{de}{DE}{drain efficiency}
\newacronym{dl}{DL}{downlink}
\newacronym{dmimo}{D-MIMO}{distributed MIMO}
\newacronym{dpd}{DPD}{digital pre-distortion}
\newacronym{dpdk}{DPDK}{Data Plane Development Kit}
\newacronym{drx}{DRX}{Discontinuous Reception Mode}
\newacronym{e2e}{E2E}{End-to-End}
\newacronym{easa}{EASA}{European Union Aviation Safety Agency}
\newacronym{ecdf}{eCDF}{empirical cumulative distribution function}
\newacronym{ecsp}{ECSP}{edge computing service point}
\newacronym{edlc}{EDLC}{electrostatic double-layer capacitors}
\newacronym{edrx}{eDRX}{Extended Discontinuous Reception Mode}
\newacronym{ee}{EE}{energy efficiency}
\newacronym{egprs}{EGPRS}{Enhanced Data Rates for GSM Evolution}
\newacronym{eirp}{EIRP}{effective isotropic radiated power}
\newacronym{em}{EM}{electromagnetic}
\newacronym{embb}{eMBB}{enhanced Mobile Broadband}
\newacronym{en}{EN}{energy neutral}
\newacronym{eol}{EoL}{End of Life}
\newacronym{epu}{EPU}{edge processing unit}
\newacronym{erp}{ERP}{effective radiated power}
\newacronym[plural=ESCs,firstplural=electronic speed controllers (ESCs)]{esc}{ESC}{electronic speed control}
\newacronym{etsi}{ETSI}{European Telecommunications Standards Institute}
\newacronym{evm}{EVM}{Error Vector Magnitude}
\newacronym{eurllc}{eRLLC}{extreme ultra reliable low latency
communication}
\newacronym{fa}{FA}{federation anchor}
\newacronym{fembb}{feMBB}{further enhanced mobile broadband}
\newacronym{fft}{FFT}{fast Fourier transform}
\newacronym{fh}{FH}{fronthaul}
\newacronym{fim}{FIM}{Fisher information matrix}
\newacronym{fom}{FoM}{figure of merit}
\newacronym{gb}{GB}{grant-based}
\newacronym{gf}{GF}{grant-free}
\newacronym{gnb}{gNB}{Next Generation Node B}
\newacronym{gnn}{GNN}{graph neural network}
\newacronym{gnss}{GNSS}{Global Navigation Satellite System}
\newacronym{gprs}{GPRS}{General Packet Radio Services}
\newacronym{gps}{GPS}{Global Positioning System}
\newacronym{gpu}{GPU}{graphics processing unit}
\newacronym{gsm}{GSM}{Global System for Mobile Communications}
\newacronym{gwp}{GWP}{Global Warming Potential}
\newacronym{harq}{HARQ}{hybrid automatic repeat request}
\newacronym{hcs}{HCS}{human-centric services}
\newacronym{hpbm}{HPBM}{half power beam width}
\newacronym{HyMPRo}{HyMPRo}{Hybrid Multi-Path Routing algorithm}
\newacronym{ib}{IB}{in-band}
\newacronym{ibo}{IBO}{input back-off}
\newacronym{ic}{IC}{integrated circuit}
\newacronym{id}{ID}{information decoding}
\newacronym{iid}{i.i.d.}{independently and identically distributed}
\newacronym{im}{IM}{intermodulation}
\newacronym{imu}{IMU}{inertial measurement unit}
\newacronym{iot}{IoT}{Internet of Things}
\newacronym{ipt}{IPT}{inductive power transfer}
\newacronym{ipy}{IPY}{Interventions per Year}
\newacronym{ir}{IR}{infrared}
\newacronym{ism}{ISM}{Industrial, Scientific and Medical}
\newacronym{isp}{ISP}{internet service provider}
\newacronym{iq}{I/Q}{in-phase/quadrature}
\newacronym{io}{IO}{input/output}
\newacronym{kpi}{KPI}{key performance indicator}
\newacronym{kvi}{KVI}{key value indicator}
\newacronym{lca}{LCA}{life cycle assessment}
\newacronym{lco}{LCO}{lithium cobalt oxide}
\newacronym{ldo}{LDO}{Low-dropout voltage regulator}
\newacronym{ldpc}{LDPC}{low-density parity-check}
\newacronym{lic}{LIC}{lithium-ion capacitor}
\newacronym{ltc}{LTC}{lithium thionyl chloride}
\newacronym{liion}{Li-ion}{lithium-ion}
\newacronym{lidar}{LiDAR}{light detection and ranging}
\newacronym{lis}{LIS}{large intelligent surface}
\newacronym{llh}{LLH}{log-likelihood}
\newacronym{lmmse}{LMMSE}{least minimum mean square error}
\newacronym{lmo}{LMO}{lithium ion manganese oxide}
\newacronym{lora}{LoRa}{Long Range}
\newacronym{lorawan}{LoRaWAN}{long-range wide-area network}
\newacronym{los}{LoS}{line-of-sight}
\newacronym{lp}{LP}{linear programming}
\newacronym{lpt}{LPT}{laser power transfer}
\newacronym{lpwa}{LPWA}{Low Power Wide Area}
\newacronym{lpwan}{LPWAN}{low-power wide-area network}
\newacronym{lrelu}{LReLU}{leaky rectified linear unit }
\newacronym{lrt}{LRT}{likelihood-ratio test}
\newacronym{ls}{LS}{least squares}
\newacronym{lsfc}{LSFC}{large-scale fading component}
\newacronym{lte}{LTE}{Long Term Evolution}
\newacronym{lto}{LTO}{lithium-titanium-oxide}
\newacronym{m2m}{M2M}{Machine to Machine}
\newacronym{mac}{MAC}{Medium Access Control}
\newacronym{mcl}{MCL}{Maximum Coupling Loss}
\newacronym{mcs}{MCS}{modulation and coding scheme}
\newacronym{mcu}{MCU}{Microcontroller Unit}
\newacronym{mec}{MEC}{multi-access edge computing}
\newacronym{mf}{MF}{matched filter}
\newacronym{mmimo}{mMIMO}{massive MIMO}
\newacronym{mimo}{MIMO}{multiple-input multiple-output}
\newacronym{miso}{MISO}{multiple-input single-output}
\newacronym{ml}{ML}{machine learning}
\newacronym[]{mlp}{MLP}{multilayer perceptron}
\newacronym{mmse}{MMSE}{minimum mean square error}
\newacronym{mmtc}{mMTC}{massive machine-typed communication}
\newacronym{mmwave}{mmWave}{millimeter wave}
\newacronym{mpc}{MPC}{multipath component}
\newacronym{mppt}{MPPT}{maximum power point tracking}
\newacronym{mr}{MR}{maximum ratio}
\newacronym{mrc}{MRC}{Magnetic Resonance Coupling}
\newacronym{mrc_em}{MRC}{maximum ratio combining}
\newacronym{mrt}{MRT}{maximum ratio transmission}
\newacronym{mse}{MSE}{mean square error}
\newacronym{mtc}{MTC}{Machine-Type Communication}
\newacronym{multi-rat}{Multi-RAT}{Multiple Radio Access Technology}
\newacronym{nb}{NB}{narrowband}
\newacronym{nbiot}{NB-IoT}{Narrowband IoT}
\newacronym{nlos}{NLoS}{non-line-of-sight}
\newacronym{nn}{NN}{neural network}
\newacronym{noma}{NOMA}{non-orthogonal multiple access}
\newacronym{nr}{NR}{New Radio}
\newacronym{oai}{OAI}{OpenAirInterface} %
\newacronym{ofdm}{OFDM}{orthogonal frequency-division multiplexing}
\newacronym{ofdmim}{OFDM-IM}{OFDM with index modulation}
\newacronym{oob}{OOB}{out-of-band}
\newacronym{ota}{OTA}{over-the-air}
\newacronym{otaa}{OTAA}{Over The Air Authentication}
\newacronym{oran}{O-RAN}{open radio-access network}
\newacronym{p2p}{P2P}{point-to-point}
\newacronym{ptp}{PTP}{precision-time protocol}
\newacronym[plural=PAs,firstplural=Power Amplifiers (PAs)]{pa}{PA}{power amplifier}
\newacronym{pae}{PAE}{power-added efficiency}
\newacronym{papr}{PAPR}{peak-to-average power ratio}
\newacronym{pc}{PC}{pilot count}
\newacronym{pcb}{PCB}{Printed Circuit Board}
\newacronym{pdcch}{PDCCH}{physical downlink control channel}
\newacronym{pdp}{PDP}{power delay profile}
\newacronym{pdsch}{PDSCH}{physical downlink shared channel}
\newacronym{per}{PER}{packet error rate}
\newacronym{pg}{PG}{path gain}
\newacronym{pl}{PL}{path loss}
\newacronym{pll}{PLL}{phase-locked loop}
\newacronym{prbs}{PRBs}{Physical Resource Blocks}
\newacronym{psd}{PSD}{power spectral density}
\newacronym{psm}{PSM}{Power Saving Mode}
\newacronym{ptw}{PTW}{Paging Time Window}
\newacronym{ptrs}{PTRS}{Phase-Tracking Reference Signals}
\newacronym{pps}{1PPS}{pulse per second}
\newacronym{qam}{QAM}{quadrature amplitude modulation}
\newacronym{qos}{QoS}{quality-of-service}
\newacronym{quadriga}{QuaDRiGa}{QUAsi Deterministic RadIo channel GenerAtor}
\newacronym{ra}{RA}{Random Access}
\newacronym{ran}{RAN}{radio access network}
\newacronym{rar}{RAR}{Random Access Response}
\newacronym{rat}{RAT}{Radio Access Technology}
\newacronym{rbs}{RBS}{radio base station}
\newacronym{re}{RE}{radio element}
\newacronym{relu}{ReLU}{rectified linear unit}
\newacronym{rllmtc}{RLLMTC}{reliable low latency machine type communication}
\newacronym{rof}{RoF}{radio-over-fiber}
\newacronym{rf}{RF}{radio frequency}
\newacronym{rfeh}{RFEH}{radio frequency energy harvesting}
\newacronym{rfid}{RFID}{radio frequency identification}
\newacronym{rfpt}{RFPT}{radio frequency power transfer}
\newacronym{fpga}{FPGA}{field-programmable gate array}
\newacronym{ris}{RIS}{reflective intelligent surface}%
\newacronym{rms}{RMS}{root-mean-square}
\newacronym{rmse}{RMSE}{root-mean-square error}
\newacronym{rrc}{RRC}{Radio Resource Connection}
\newacronym{rreq}{RREQ}{route request packet}
\newacronym{rsrp}{RSRP}{Reference Signals Received Power}
\newacronym{rss}{RSS}{received signal strength}
\newacronym{rssi}{RSSI}{received signal strength indicator}
\newacronym{rtc}{RTC}{Real Time Clock}
\newacronym{rtk}{RTK}{Real Time Kinematics}
\newacronym{rts}{RTS}{ray tracing simulator}
\newacronym{rw}{RW}{RadioWeaves}
\newacronym{rx}{RX}{receiver}
\newacronym{rzf}{RZF}{regularized zero forcing}
\newacronym{duc}{DUC}{digital up-converter}
\newacronym{sdm}{SDM}{sigma-delta modulator}
\newacronym{poe}{PoE}{power-over-Ethernet}
\newacronym{sa}{SA}{synchronization anchor}
\newacronym{scfdma}{SCFDMA}{Single-Carrier Frequency Division Multiple Access}
\newacronym{sdg}{SDG}{Sustainable Development Goal}
\newacronym{sdn}{SDN}{software-defined network}
\newacronym{sdr}{SDR}{software-defined radio}
\newacronym{sf}{SF}{Spreading Factor}
\newacronym{sfn}{SFN}{single frequency network}
\newacronym{sinr}{SINR}{signal-to-interference-plus-noise ratio}
\newacronym{siso}{SISO}{single-input single-output}
\newacronym{slam}{SLAM}{simultaneous localization and mapping}
\newacronym{slc}{SLC}{spatial leakage suppression}
\newacronym{smc}{SMC}{specular multipath component}
\newacronym{smps}{SMPS}{Switched Mode Power Supply}
\newacronym{sndr}{SNDR}{signal-to-noise-and-distortion ratio}
\newacronym{snidr}{SNIDR}{signal-to-noise-and-interference-and-distortion ratio}
\newacronym{snr}{SNR}{signal-to-noise ratio}
\newacronym{soc}{SoC}{state of charge}
\newacronym{ssb}{SSB}{synchronisation signal block}
\newacronym{steam}{STEAM}{science, technology, engineering, the arts, and mathematics}
\newacronym{svd}{SVD}{singular value decomposition}
\newacronym{sdof}{SDoF}{sigma-delta over fiber}
\newacronym{tau}{TAU}{Tracking Area Update}
\newacronym{tdd}{TDD}{time division duplexing}
\newacronym{tdoa}{TDOA}{time-difference-of-arrival}
\newacronym{toa}{TOA}{time-of-arrival}
\newacronym{trl}{TRL}{technology readyness level}
\newacronym{trp}{TRP}{Transmission Reception Point}
\newacronym{ttm}{TTM}{Time To Market}
\newacronym{ttn}{TTN}{The Things Network}
\newacronym{tx}{TX}{transmitter}
\newacronym{tcer}{TCER}{transported to consumed energy ratio}
\newacronym{uav}{UAV}{unmanned aerial vehicle}
\newacronym{udp}{UDP}{User Datagram Protocol}
\newacronym{ue}{UE}{user equipment}
\newacronym{ugv}{UGV}{unmanned ground vehicle}
\newacronym{uhf}{UHF}{ultra high frequency}
\newacronym{ul}{UL}{uplink}
\newacronym{ula}{ULA}{uniform linear array}
\newacronym{upa}{UPA}{uniform planar array}
\newacronym{ura}{URA}{uniform rectangular array}
\newacronym{urllc}{URLLC}{Ultra-Reliable Low Latency Communications}
\newacronym{ummtc}{umMTC}{ultra massive machine type communication}
\newacronym{uv}{UV}{unmanned vehicle}
\newacronym{uwb}{UWB}{ultrawideband}
\newacronym{usrp}{USRP}{Universal Software Radio Peripheral}
\newacronym{wb}{WB}{wideband}
\newacronym{wpt}{WPT}{wireless power transfer}
\newacronym{wrsn}{WRSN}{Wireless Rechargeable Sensor Network}
\newacronym{wsn}{WSN}{Wireless Sensor Network}
\newacronym{wr}{WR}{White Rabbit}
\newacronym{z3ro}{Z3RO}{zero third-order distortion}
\newacronym{zf}{ZF}{zero-forcing}
\newacronym{zmcscg}{ZMCSCG}{zero mean circularly symmetric complex Gaussian}

\usepackage[inline]{enumitem}
\usepackage{booktabs}
\usepackage{url}
\usepackage[font={small}]{caption}
\usepackage{subcaption}
\usepackage[capitalise]{cleveref}
\usepackage{siunitx}
\usepackage{multirow} 

\usepackage{array}
\newcolumntype{H}{>{\setbox0=\hbox\bgroup}c<{\egroup}@{}}

\usepackage[dvipsnames]{xcolor}

\usepackage{xspace}
\newcommand{\sigdel}{$\Sigma\Delta$\xspace}

\usepackage{tikz}

\newcommand{\tikzcmark}{%
\tikz[scale=0.23] {
    \draw[line width=0.7,line cap=round] (0.25,0) to [bend left=10] (1,1);
    \draw[line width=0.8,line cap=round] (0,0.35) to [bend right=1] (0.23,0);
}}
\newcommand{\yes}{\protect\tikzcmark}
\newcommand{\no}{} %

\usepackage[backend=biber, style=ieee, citestyle=numeric-comp]{biblatex} %
\AtBeginBibliography{\footnotesize}

\newcommand{\update}[1]{#1\xspace} %

\begin{document}

\maketitle
\let\thefootnote\relax\footnotetext{%
This manuscript has been accepted for MTT-S TC-23 Wireless Communications Focus Issue of the IEEE Microwave Magazine.\\
\begin{itemize*}
    \item[] \textbf{Gilles Callebaut}, Department of Electrical Engineering, KU Leuven, BE-9000 Ghent, Belgium, Email: gilles.callebaut@kuleuven.be, ORCID: 0000-0003-2413-986X.
    \item[] \textbf{Liang Liu}, Department of Electrical and Information Technology, Lund University, SE-22100 Lund, Sweden, Email: liang.liu@eit.lth.se, ORCID: 0000-0001-9491-8821.
    \item[] \textbf{Thomas Eriksson}, Department of Electrical Engineering, Chalmers University of Technology, SE-41296 Göteborg, Sweden, Email: thomase@chalmers.se, ORCID: 0000-0002-2087-7227.
    \item[] \textbf{Liesbet Van der Perre}, Department of Electrical Engineering, KU Leuven, BE-9000 Ghent, Belgium, Email: liesbet.vanderperre@kuleuven.be, ORCID: 0000-0002-9158-9628.
    \item[] \textbf{Ove Edfors}, Department of Electrical and Information Technology, Lund University, SE-22100 Lund, Ove.Edfors@eit.lth.se, ORCID 0000-0001-5966-8468.
    \item[] \textbf{Christian Fager}, Department of Microtechnology and Nanoscience, Chalmers University of Technology, SE-41296 Göteborg, Sweden, Email: christian.fager@chalmers.se, ORCID: 0000-0001-8228-0736.
\end{itemize*}}

\vspace{-1.6cm}%

\begin{figure*}[ht!]
    \centering
\includegraphics[width=0.75\linewidth]{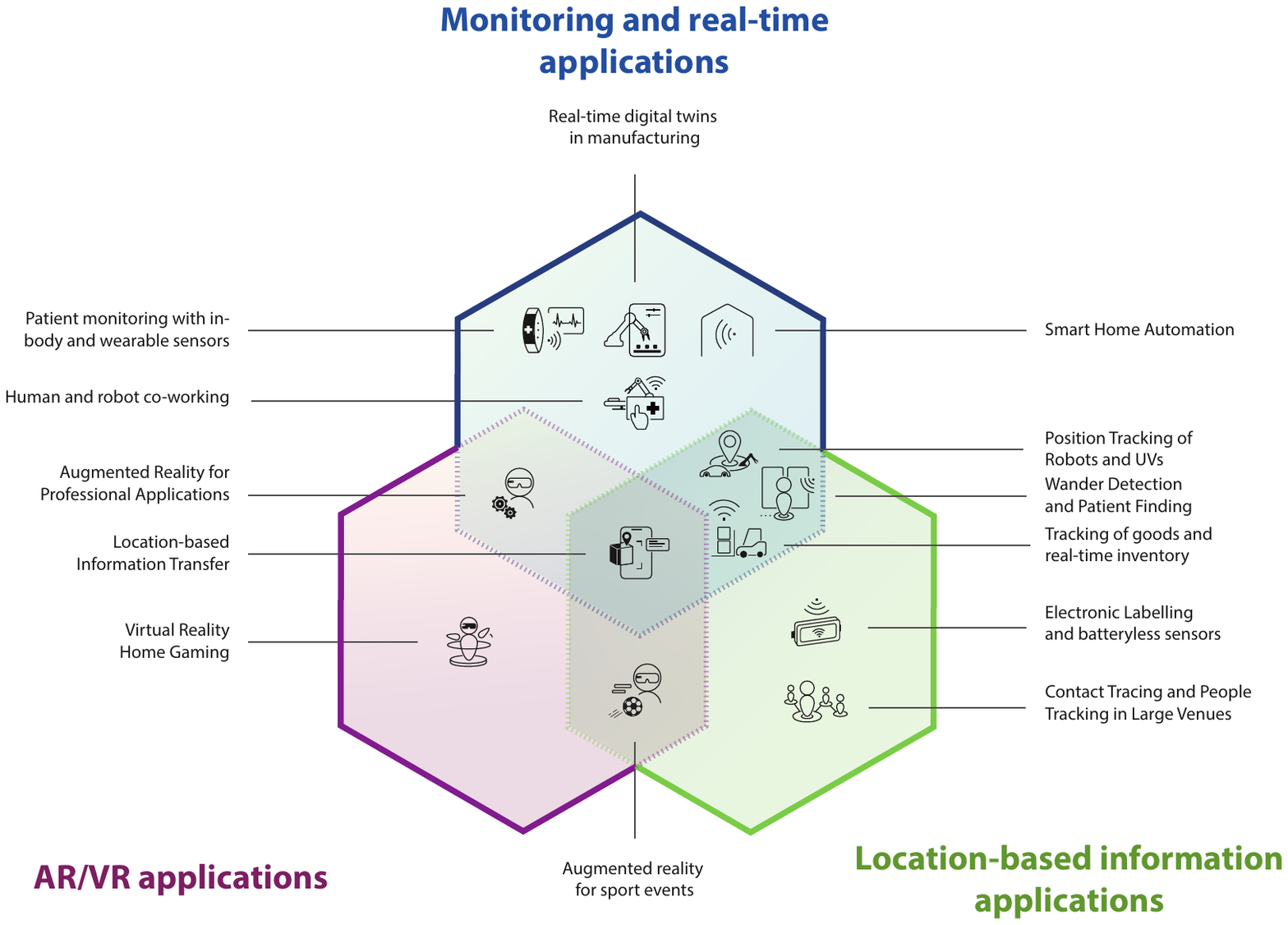}
    \caption{Envisioned interactive use cases in need of \gls{6g} (adopted with permissions from~\cite{REINDEERD1.1}).}
    \label{fig:usecases}
  \end{figure*}

\section{Why \gls{6g} is wished for and needs new testbeds}\label{sec:motivation}%

The proof of the pudding lies in eating, that is why \gls{6g} testbeds are essential in the progress towards the next generation of wireless networks. Theoretical research towards \gls{6g} wireless networks proposes advanced technologies to serve new applications and drastically improve the energy performance of the network. Testbeds are indispensable to validate these new technologies under more realistic conditions. This paper clarifies the requirements for \gls{6g} radio testbeds, reveals trends, and introduces approaches toward their development.

The \acrlong{6g} of mobile networks is expected to be introduced starting 2030. Research and development toward this novel generation of networks started several years ago in different regions of the world, e.g.,~\cite{9625032}. A myriad of applications and novel use cases, ranging from \gls{ar} and mixed reality (MR) in professional and entertainment contexts to support autonomous systems and connecting a massive number of battery-less devices, are envisioned to be served by these networks~\cite{OnTheRoadTo6G}. The analysis of these use cases has led to a harmonized view of which functions and capabilities \gls{6g} networks should offer. Summarizing, the desired features for \gls{6g} can be classified into three main categories:
\begin{enumerate}
    \item \textit{Improving diverse wireless network services}: delivering higher throughput and network capacity, ultra-reliable operation, imperceptible latency, and connectivity to many devices with an extremely low-energy budget.
    \item \textit{Providing novel functionalities}: precise positioning and sensing~\cite{9569364}, and communication with energy-neutral devices requiring wireless-power transfer~\cite{7984754}.
    \item \textit{Enhancing coverage}: delivering more uniform and truly ubiquitous services.
\end{enumerate}
\update{In addition} to new features, \update{the intent} for \gls{6g} has been expressed to improve energy efficiency \update{by} several orders of magnitude with respect to previous generation technologies, in order to keep up with the increasing volumes of mobile data without an associated increase in carbon footprint. More holistic value-related targets are being discussed regarding the dual notions of a sustainable \gls{6g}, and \gls{6g} for sustainability. 

To support the above, novel technologies need to be developed for \gls{6g} networks \cite{gustavsson2021implementation}. Candidate technologies receiving focus from the academic and industrial community are:
\begin{itemize}
    \item \Gls{cf} \gls{mmimo} deploying a distributed infrastructure and applying spatial multiplexing to offer uniform good service levels and energy-efficient operation for large amounts of users~\cite{https://doi.org/10.48550/arxiv.2208.14048, interdonato2019ubiquitous, Palacios2020, terminology}.
    \item Operation at higher frequencies~\cite{gunnarsson2021mmwave, 7534832, 7967756, Moerman2022}, in particular up to sub-THz, where the large available bandwidth enables high throughput and precise positioning.
    \item The use of \gls{ai} technologies to increase performance and cope with the increased complexity in the networks~\cite{9446676}.
    \item \Gls{ris} as a technology~\cite{9998527}, to contribute to improving consistent coverage, which is especially challenging at \gls{mmwave} frequencies and above.
    \item Integration of terrestrial and non-terrestrial networks~\cite{9275613}.
\end{itemize}

Experiments in testbeds can complement and strengthen theoretical work by providing validation under \textit{more realistic conditions}, and hence increase the confidence level in these newly introduced technologies. Moreover, and at least as important, quite often these experiments reveal limitations or raise \update{doubt about} theoretical models or assumptions made. To test these assumptions and theoretical models under \textit{more realistic conditions}, we require space, time, and actual hardware in order to represent 'real-life' deployment scenarios. 
Therefore, we \update{use this} definition for a wireless testbed: ``an experimental environment in which tests can be performed over real propagation channels, using real hardware and possibly operating in real-time''.  

Innovative transmission schemes \update{need to be} developed to a higher \gls{trl} for testbed implementation. In this process, it is possible to refine models to better represent reality. Also, it allows developing models for and assess the sensitivity to hardware impairments. At the same time, testbeds typically provide much more flexibility and openness, potentially both in terms of hardware and data, than commercial systems. From that perspective, testbeds also contribute to a better technological understanding for the broad R\&D community. %

As the technological progress towards \gls{6g} has been intensifying, it is essential to also develop new testbeds for \gls{6g} networks for two main reasons: (i) to validate new technologies that were not present in previous generations raises new requirements on the testbeds to validate these,  and (ii) the development of testbeds in itself often presents a first attempt and preview on challenges that will be encountered in real deployments.

\subsection{Scope}
 
We focus on the validation of the new radio access, i.e., wireless, solutions to provide widely available high-performance services, in terms of capacity and reliability. Typical interactive applications to be served in dense interconnected environments range from real-time monitoring and \gls{ar}/\gls{vr} to location-based information applications, as illustrated in~\cref{fig:usecases}.  
The key candidate technology to provide this performance is expected to be extreme \gls{dmimo}, operating in frequency bands below \SI{20}{\giga\hertz}.  
As a first contribution, this paper provides an overview of the corresponding \gls{6g} testbed requirements, based on the expected technological innovation that will be at the foundation of \gls{6g} networks and new functionalities to be supported. Further, it explains several complementary approaches to construct and operate scalable \gls{6g} testbeds to validate diverse functions and capabilities. Potential approaches to achieve synchronization are \update{detailed}, as experience in actually building testbeds \update{shows} that this poses a main, if not the main, challenge for \gls{dmimo}. 
Targets for \gls{6g} testbeds are broad, as the envisioned `network of networks' will \update{among others} include both terrestrial and non-terrestrial components and serve applications with frequencies up to sub-THz~\cite{OnTheRoadTo6G}. While the latter paper clarifies \gls{6g} requirements, we here focus on how to attain these are addressed in the testbeds.  All anticipated propagation scenarios and frequency bands for \gls{6g}~\cite{wang2022pervasive} \update{can not be} validated in one testbed. The approaches presented in this paper are targeted to the radio access technologies that are expected to become the main capacity bearer for the so-named extreme experience services~\cite{9625032}. Complementary testbed developments focus on operation in very high frequency bands~\cite{zhu2023ultra} and~\cite{SEN2020107370}, or on higher layer networking technologies, e.g., \gls{sdn}~\cite{9785750,10.1145/3286680.3286683}.

\subsection{Different levels of testbeds}
\begin{figure*}
    \centering
    \includegraphics[width=0.55\textwidth]{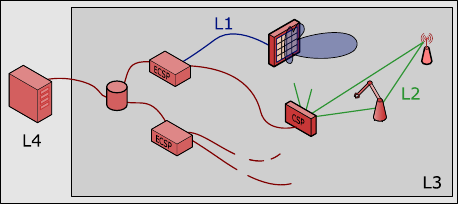}
    \caption{Overview of a \gls{6g} radio system with different testbed scopes. At level~1 (L1), new hardware and configurations are evaluated. Level~2 (L2), i.e., channel sounders, study the novel propagation conditions in \gls{6g} scenarios and frequency bands. Level~3 (L3) and~4 (L4) are designed to test, e.g., distributed processing in real-life and real-time conditions, respectively.}\label{fig:testbed-levels}
\end{figure*}
As discussed in the previous section, \gls{6g} presents a myriad of new research challenges, each having \update{different scales, aims,} and performance metrics. 
In order to discuss the different testbeds, we distinguish them based on their scope and intention. An overview of these different levels and a common architecture is presented in~\cref{fig:testbed-levels}. Following the terminology introduced in~\cite{terminology}, a \gls{6g} system architecture could exist out of several connected \glspl{ecsp}, which coordinate one or multiple \glspl{csp} and perform joint processing. The term \gls{csp} was introduced to indicate that the conventional \gls{ap} is expected to support more than just communication in \gls{6g}, e.g., \gls{wpt}. This \update{implies} that these \glspl{csp} will host a variety of hardware such as radio, charging, processing, data storage and other sensing elements. We categorized four levels of testbeds based on their capabilities and intended use, as illustrated in~\cref{fig:testbed-levels} and elaborated below.

\paragraph*{Level~1 Testbed - \Gls{p2p} Link Evaluation}
Moving to higher frequencies requires studying \gls{p2p} link operation to test new hardware architectures. For instance, \citeauthor{9720486}~\cite{9720486} present a reconfigurable tile-based antenna array system for \gls{mmwave} frequencies. It enables studying different \gls{bf} architectures such as analog and hybrid \gls{bf}, antenna positioning and layout at both the \gls{bs} and \gls{ue}. Furthermore, given that users will be potentially in near-field in \gls{dmimo}, novel wavefront designs need to be developed and tested in real scenarios, e.g., Bessel beamforming~\cite{10052090}. \Acrlong{p2p} sub-THz communication experiments \update{also belong} to this category. Additionally, new fronthaul media for connecting the distributed \gls{csp} and \gls{ecsp} nodes are being investigated. Analog- and various digital \gls{rof} techniques are expected to play an important role~\cite{rommel2020towards,breyne2017comparison,Bao2023}. These two examples of L1 testbeds are depicted in blue in~\cref{fig:testbed-levels}.

\paragraph*{Level~2 Testbed - Channel Sounders}
To study the new propagation conditions as an effect of distributing resources, moving to higher carrier frequencies~\cite{wang2022pervasive} and having an increased number of antenna elements with respect to conventional \gls{mmimo}, channel sounders are used. In essence, a channel sounder consists of one or more \gls{tx} and \gls{rx} chains, where the transmitted signals are recorded at the \gls{rx}, and processed to investigate the propagation channel, as illustrated in green in~\cref{fig:testbed-levels}. Different approaches are used to sample the channel. Virtual arrays can be created by moving a single antenna to multiple locations. Another approach is to have multiple antennas connected to one \gls{rf} chain, where the antennas are time-multiplexed via an \gls{rf}~switch. For example, in~\cite{7534832}, \citeauthor{7534832} study the \gls{pl} models, \gls{cir}, coherence time, and more, in different environments and wall materials with a \gls{rf}-switched 16-antenna channel sounder operating at \SI{83.5}{\giga\hertz}.

\paragraph*{Level~3 and~4 Testbeds}
In contrast to the channel sounders, real-life (L3) testbeds do not require fine-grained propagation channel knowledge, such as for example \gls{aoa}, but are designed to investigate end-to-end communication performance, as for instance the network's energy-efficiency. In L3 testbeds, real-life measurements are conducted and processed in an offline stage. L4 testbeds extend L3 by supporting real-time processing of signals to evaluate algorithms requiring real-time interaction between the network and the \glspl{ue}, unable to be performed offline. This is further discussed in \cref{sec:offreal}. The real-time operation (L4), in contrast to L3 testbeds, is highlighted by the use of a real-time processing infrastructure in~\cref{fig:testbed-levels}.

\section{Moving from 5G to 6G -- Challenges and Requirements for 6G Testbeds}\label{sec:requirements} %
Here, the new requirements for \gls{6g} testbeds with respect to previous generations are discussed. As motivated in~\cref{sec:motivation}, \gls{6g} testbeds are required to enable research evaluating new technologies and architectures, therefore our focus will be on testbeds targeting Level 3 and 4. Specifically, the move towards a distributed deployment in 6G, necessitates a number of new methods to cope with, including i) the geographical dispersion of processing resources and ii) the high number of \glspl{csp}. On top of this, providing real-time operation of a \gls{dmimo} testbed imposes new challenges.

\subsection{Offline vs. real-time signal processing}%
\label{sec:offreal}
Both offline and real-time signal processing are important features for a \gls{6g} testbed. Offline processing requires the capability of recording data for a certain time interval for later analysis, giving the flexibility to develop and evaluate a wide range of \update{signal-processing} algorithms. On the other \update{hand}, real-time processing enables system measurement and testing in fast-changing environments. The real-time (L4) testbed can also serve as the system-level tester for hardware solutions of specific function blocks (e.g., digital signal processors, data converters, and transceivers).

The first challenge for real-time signal processing is the high throughput requirement. For instance, depending on the selected algorithms, the required processing capability to support real-time multi-user detection can grow exponentially with the number of \gls{bs} antennas and the number of simultaneously served users.  
Processing latency is another challenge for real-time testbeds, especially for time-critical use cases and/or in fast-changing environments. There are two processing latency constraints when developing a real-time testbed. The first one is the end-to-end latency, e.g., \gls{tx}-to-\gls{rx}, which is bounded by application requirements. For applications such as real-time digital twins in manufacturing, and human and robot co-work, the end-to-end latency should be around \SI{1}{\milli\second}~\cite{truskaller2021use}. The second latency constraint relates to \gls{tdd} switching time in reciprocity-based \ac{mmimo}. The process of channel estimation using uplink pilots, calculating precoding coefficients, and transmitting the precoded data all need to be finished within the time interval between the uplink pilot and downlink data in a radio frame~\cite{7931558}. This time interval can be in the sub-millisecond range, depending on the supported coherence time and thus the radio frame structure. For instance, the processing latency constraint is around \SI{0.4}{\milli\second} to support \SI{70}{\kilo\meter\per\hour} mobility at \SI{3.7}{\giga\hertz}~\cite{7931558}. The latency requirement is even more critical for decentralized processing architectures, where the data transfer between processing units may take a substantial amount of time.

\subsection{Moving to distributed processing architectures}\label{sec:co_vs_d_mimo}

\begin{figure*}[tb!]
  \centering
  \begin{subfigure}[b]{0.48\linewidth}
    \centering
    \includegraphics[height=3.5cm]{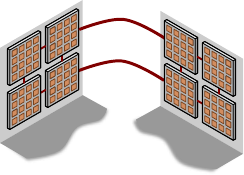}
    \caption{Decentralized antenna arrays with distributed processing resources in each array.}
  \end{subfigure}\hfill%
  \begin{subfigure}[b]{0.48\linewidth}
    \centering
    \includegraphics[height=3cm]{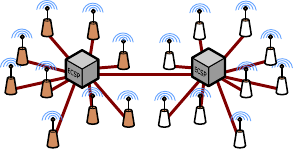}
    \caption{Fully distributed antennas connecting to \glspl{ecsp} for centralized processing.}
  \end{subfigure}
  \caption{Distributed and centralized processing architectures applied in \ac{mmimo} systems with different flavors of antenna distributions.}%
  \label{fig:disproc}
\end{figure*}

Independent of the physical locations of the antennas (co-located or distributed), there are various processing strategies that can be applied, affecting the data transfer between processing units in different ways. Here we will focus on the two main processing strategies, namely centralized and decentralized processing~\cite{sanchez2020decentralized,sanchez2022distributed}, as illustrated in~\cref{fig:disproc}. The centralized solution faces critical implementation challenges when up-scaling the number of antennas and the signal bandwidth, especially considering the aggregated data rate to the central processing unit. For a \gls{6g} \ac{mmimo} system with \num{256}~\ac{bs} antennas, 25 single-antenna users, and \SI{100}{MHz} bandwidth, the aggregated data rate reaches around \SI{750}{Gb/s}, exceeding the data-rate limit of commonly available interconnection solutions. The aggregated data may be distributed over different physical links, however, the number of \gls{io} connections is generally limited on a practical processing hardware platform. Next to dedicated interconnects, a candidate medium to convey the data is Ethernet, where software is being developed, e.g., \gls{dpdk}~\cite{9183752}, to enable high-throughput real-time data transfer.

Decentralized processing (including distributed processing algorithms and the corresponding distributed architectures) enables closer-to-antenna processing and avoids extensive data aggregation, and thus is more suited for scalable implementations. This method follows more closely \update{a} generic architecture, using \glspl{ecsp} and \glspl{csp} as depicted in~\cref{fig:testbed-levels}. For instance, the recursive least square algorithm is a decentralized implementation of the \gls{zf} detection, which is mapped to a Daisy-chain topology \update{distributing the data only} between neighboring processing units in the chain~\cite{sanchez2020decentralized}. This, however, implies long processing latency to propagate the signal processing serially through the entire chain. This could be partially overcome by applying more parallel topologies, e.g., mesh and tree~\cite{sanchez2022distributed}. In physically distributed \ac{mmimo} systems, it is expected that numerous antennas are available around (and close to) the users, i.e., \glspl{ecsp}. Algorithms with local-only processing, e.g., \ac{mrt}, are likely to provide good enough system performance and approach the centralized \gls{zf} performance. The corresponding data transfer between processing units can then be minimized.

Data storage is another implementation challenge in the centralized solution. The required memory size to store the channel matrices of the \SI{100}{MHz} $256\times25$ \ac{mmimo} system is \SI{380}{Mb} (24-bits per channel matrix element). In the distributed processing architecture, data storage can be distributed over many processing units, making it possible to use more cost and energy-efficient on-chip memory solutions for each processing unit.

\subsection{Calibration and Synchronisation}\label{sec:calibrate_sync}

\begin{table*}[hbtp] 
    \centering
    \caption{Calibration and Synchronization mechanisms employed in \gls{6g} testbeds. Addressing the calibration issue is indicated by \yes. How well the mechanism can scale with both the geographical dispersion and the number of \glspl{csp} in \gls{dmimo} is marked by, $\bullet\circ\circ$, $\bullet\bullet\circ$ and $\bullet\bullet\bullet$. More details regarding the mechanisms can be found in the included references (Ref. column). %
    }\label{tab:sync}
    \begin{tabular}{p{5cm} H l l ll l l l H l}
    \toprule
         Mechanism & Description & \multicolumn{2}{c}{Frequency} & Phase & Time & \multirow{2}{*}{\Gls{dmimo}} & \multicolumn{2}{c}{Carrier} & Complexity & \multirow{2}{*}{Ref.}\\
         && Drift & Offset & Offset & Offset & &Sub-10GHz & \gls{mmwave}\\ \midrule
         RF and BB PLL sharing & & \yes & \yes & \yes & \yes & \no & \yes & \no & High\\
         Reference clock (e.g., \SI{10}{\mega\hertz} and \acrshort{pps}) & eg via \gls{gnss} or dedicated cabling & \yes & \yes & \no & \yes & $\bullet\circ\circ$ & \yes & \yes & Moderate & \cite{techtile,7534832, POWDER, 7534832}\\
         Ethernet-based\\
         \hspace{0.5cm} White Rabbit & & \yes & \yes & \yes & \yes & $\bullet\bullet\bullet$ & \yes & \yes & Low & \cite{9915020, POWDER}\\
         \hspace{0.5cm} \Gls{ptp} & & \yes & \yes & \no & \yes & $\bullet\bullet\bullet$ & \yes & \yes & Low & \cite{CFRAN, techtile, 9183752}\\
         \Acrlong{ota}\\
         \hspace{0.5cm} UE-assisted sync & & \yes & \yes & \yes & \yes & $\bullet\bullet\bullet$ & \yes & \yes & Low &  \cite{rahman2012fully,rogalin_scalable_2014} \\
         \hspace{0.5cm} Beacon/anchor nodes & & \yes & \yes & \yes & \yes & $\bullet\bullet\circ$ & \yes & \yes & Low & \cite{Balan2013,Rahman2012} \\
         \Acrlong{rof} &  &  &  &  &  &  &  &  & & \\
         \hspace{0.5cm} Analog & \yes & \yes & \yes & \yes & \yes & $\bullet\bullet\circ$ & \yes & \yes & & \cite{rommel2020towards,Moerman2022}\\
         \hspace{0.5cm} Digital (\sigdel) & \yes & \yes & \yes & \yes & \yes & $\bullet\bullet\circ$ & \yes & \yes & & \cite{Pessoa2014,wu2019distributed,Sezgin2018}\\
         \bottomrule
    \end{tabular}
\end{table*}

  A main challenge in most wireless systems in general, and in testbeds in particular, is \textit{calibration} and \textit{synchronization}. Here we use the word calibration in a general sense, including gain and phase calibration, linearization, \gls{iq} imbalance compensation, and impairment mitigation in general. Synchronization includes mismatch and drift related to the oscillators, e.g., phase, sampling time, and carrier frequency calibration. In a testbed, part of the calibration and synchronization can be done through offline procedures, using lab instruments. The calibration needs related to hardware are generally slow-varying compared to the rapid changes in the propagation channel, such that calibration procedures can be applied fairly infrequently~\cite{Palacios2020}. However, it is usually necessary to re-calibrate the testbed when it is used, preferably only using the hardware in the testbed itself, potentially based on dedicated signals. In many cases, it is also desirable to study calibration itself, since it is an integral part of the achievable signal quality, i.e., through an L1 testbed. Below, we discuss approaches that have been proposed in literature, useful for calibration of a communication network, yet also possible to implement in a testbed for research.
  
\paragraph*{Calibration} %

To calibrate a transmitter, it is necessary to capture the transmitted signal to enable later characterization and compensation, which is historically often achieved by observation receivers at the transmitter outputs. However, in a multi-antenna transmitter, the overall transmitted signal is a combination of all the antenna signals, and the output of individual transmitters is not representative of what the receivers will see. Further, with the development towards integrated systems, accessing the signals at the transmitter outputs can be difficult. These challenges require \gls{ota} signal acquisition for calibration, generally through one or more external receivers.

With phased-array \acrlong{bf}, needing knowledge of beam directions, absolute calibration is \update{required}, i.e., the transmitters and receivers are each calibrated to reach the desired linear operation as accurately as possible~\cite{Bourdoux2003}. In contrast, for reciprocity-based communication, as in \gls{mmimo}, it is enough to ensure that \ac{tx} and \ac{rx} \ac{rf} chains are reciprocal~\cite{Kaltenberger2010} 
using \emph{reciprocity calibration}. This calibration can be done in collaboration with the \glspl{ue} by pilot signalling~\cite{Kaltenberger2010}, but it is preferable to avoid this. ~\citeauthor{Shepard2012}~\cite{Shepard2012} and \citeauthor{rogalin_scalable_2014}~\cite{rogalin_scalable_2014} have proposed to instead exchange pilots only between the \glspl{csp} by, possibly also with an additional external reference antenna. Alternatively, \citeauthor{Vieira2014}~\cite{Vieira2014} show that pilot-free calibration can be performed by exploiting antenna coupling. 
\Gls{ota} linearization (i.e., compensation of nonlinear amplifiers) in a \gls{mimo} setting is studied in, e.g.,~\cite{Braithwaite2020}. %
The calibration challenge in \ac{dmimo} is similar to the challenge in co-located \ac{mimo}. However, the issue of synchronization is more challenging for distributed systems; we discuss this below.

\paragraph*{Synchronization}
Synchronization, i.e., calibration connected to oscillators, is a crucial topic in communication networks. \Cref{tab:sync} summarizes methods for synchronization and indicates whether they are suited for co-located or distributed systems, as well as applicable to sub-\SI{10}{\giga\hertz} or \gls{mmwave} carrier frequencies. \citeauthor{Mudumbai2009}~\cite{Mudumbai2009} and \citeauthor{Nasir2016}~\cite{Nasir2016} also provide an overview of timing and carrier synchronization algorithms, for co-located and distributed MIMO systems. 
In a multi-user \ac{mimo} system, the \glspl{ue} have their own independent oscillator, which poses multiple synchronization tasks for each \ac{csp}. With co-located \ac{mimo}, the oscillators or \glspl{pll} can be shared among the antennas. However, in a \ac{dmimo} system, this mechanism does not scale well with increased geographical distance between the \glspl{csp}.
The synchronization is one of the critical issues in \ac{dmimo}, and it is therefore of high priority to study in \gls{6g} testbeds.

\begin{figure*}
   \centering
   \begin{subfigure}[b]{0.5\linewidth}
    \centering
   \includegraphics[width=0.85\linewidth]{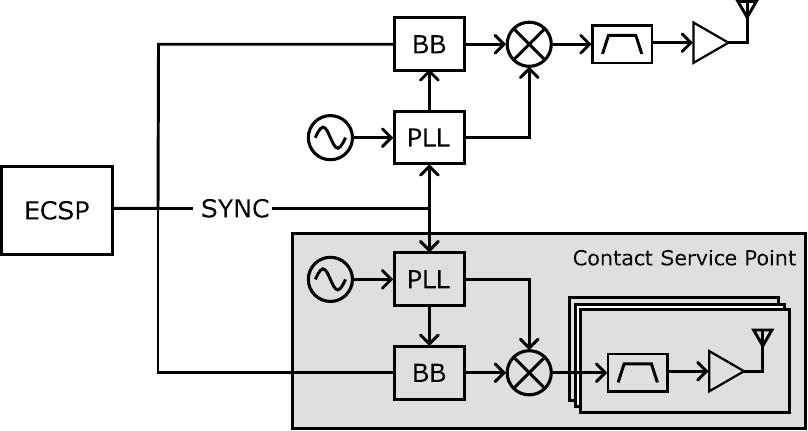}
   \caption{}\label{fig:example-sync-1}
   \end{subfigure}\hfill%
   \begin{subfigure}[b]{0.5\linewidth}
    \centering
   \includegraphics[width=0.95\linewidth]{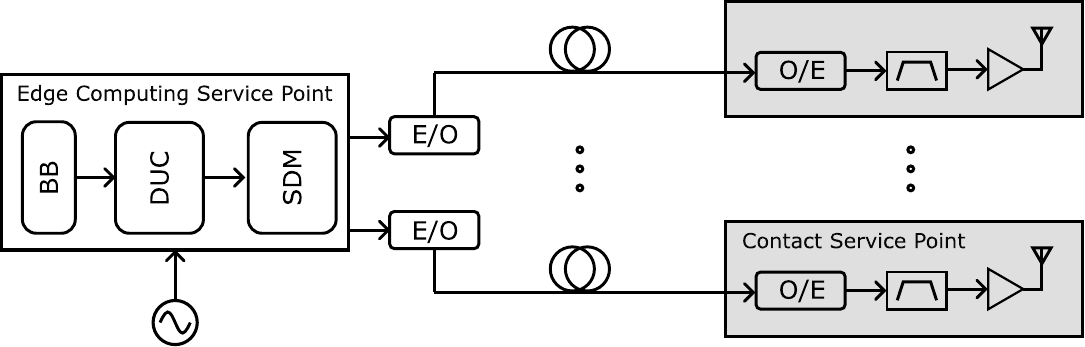}
   \caption{}\label{fig:example-sync-2}
   \end{subfigure}
   \caption{%
   Synchronization mechanisms between distributed \glspl{csp} using (a) a sync signals and (b) \gls{sdof}. The sync signal can be either an Ethernet-based or dedicated cabling or \gls{ota} technique, the principle remains that instead of sharing a \gls{lo}, the \gls{pll} is calibrated based on an external sync signal. Used abbreviations: \acrfull{bb}, \acrfull{duc}, \acrfull{sdm}, optic-to-electrical (O/E).}%
   \label{fig:example-sync}
\end{figure*}

Beamforming relies on multiple antennas, operating in a phase-synchronous way, such that the receivers experience constructive addition of multiple antenna signals. To achieve such phase-coherent \ac{dl} transmission, or \gls{ul} reception, carrier frequency, sample time and phase must be accurately synchronized. When the signals are not phase-aligned at the user, they no longer constructively interfere, degrading the signal power and quality. 
The techniques to obtain coherency can be divided into connected synchronization, i.e., using cables or optical fibers, or wireless \gls{ota} synchronization.  
Connected synchronization can be performed by sharing the \gls{rf} oscillator between the radios. As mentioned, this is problematic for high frequencies and when the distances to remote antennas increases. An alternative is to share a low-frequency reference clock (e.g., at around \SI{10}{MHz}), which can be combined with a \gls{pps} signal. This approach is also used in the Techtile testbed,~\cref{sec:Techtile}. Another possibility, used in the \gls{mate} testbed~\cref{sec:MATE_testbed}, is to distribute an \gls{if} clock which is frequency-multiplied at the antennas to the \gls{rf} frequency.
To improve the scalability further, Ethernet-based approaches can be implemented, such as \gls{wr}~\cite{Bigler2018}. This is illustrated in~\cref{fig:example-sync-1}. 
To achieve a synchronized and phase coherent \ac{mimo} \ac{dl}, an all-digital \gls{rof} solution using bandpass \sigdel~coding has been proposed~\cite{Pessoa2014}, depicted in~\cref{fig:example-sync-2}. Compared to analog \gls{rof} solutions, it enables a relatively simple and low-cost and low-complex implementation \glspl{csp}~\cite{Sezgin2018,wu2019distributed,breyne2017comparison}. A \gls{dmimo} testbed based on this approach is presented in~\cref{sec:DRoF_testbed}.  
There are also several \gls{ota} techniques proposed in the literature. Carrier frequency, sample time and phase can be estimated and compensated, using pilots shared with the \glspl{ue}~\cite{rogalin_scalable_2014} or feedback signalling~\cite{rahman2012fully,quitin2012demonstrating}.
Some proposals in the literature assume a beacon signal transmitted by a central node~\cite{Balan2013}, 
potentially leading to problems when the number of \glspl{csp} are scaled up. Decentralized schemes, e.g., based on multiple anchor nodes, are studied in, e.g.,~\cite{Rahman2012,CFRAN}.

\subsection{UE data collection}
To acquire the necessary signals at the \gls{ue}, different platforms can be used. The primary function of such platforms is to navigate the \gls{ue} data collection device to a predetermined position. For indoor environments, different rovers are deployed, often able to navigate in a 2D space~\cite{6418865, alkhateeb2022deepsense, emulab, LuViRA, Bao21}. Others~\cite{8757118, techtile} have adopted, e.g., a scissor lift, to support automated 3D sampling. The position of the data collector platform is determined via \gls{slam}~\cite{8757118}, \gls{lidar}~\cite{7534832}, cameras~\cite{emulab, 6418865, LuViRA}, fixed marking~\cite{techtile, Bao21}, mechanical positioners, e.g., CNC XY-table, or a combination of these techniques. In outdoor scenarios, drones~\cite{Ponce2021, alkhateeb2022deepsense} or humans (pedestrians~\cite{alkhateeb2022deepsense}) are used to move the \glspl{ue}. Next to using drones to emulate \glspl{ue} positions, \citeauthor{9771649}~\cite{9771649}, use drones to represent multiple \glspl{ap} in a \gls{cf} system. \Gls{gnss} is often employed to acquire the position~\cite{Ponce2021}, sometimes extended with \gls{rtk}~\cite{alkhateeb2022deepsense} to improve the location accuracy. As elaborated in~\cite{alkhateeb2022deepsense}, we also advocate \update{acquiring} multi-modal sensor measurements during data collection, such as weather conditions, images, and 3D~\gls{lidar} to support studying vision-aided algorithms for, e.g., beam and blockage prediction in \gls{mmwave} systems~\cite{charan2021vision}.

\subsection{Open Testbeds}
To facilitate widespread adoption and interoperability, \gls{6g} systems require open standards, interfaces, data sets, hardware, and software. We, here, highlight some endeavors made towards open \gls{6g} testbeds and networks. 

\subsubsection{Open Interfaces, Non-Proprietary Hardware and Software}
While not the focus of this work, \gls{oran}, next to others, has emerged due to the need for interoperability and openness in next-generation networks. The \update{emergence of open systems} indicates a shift towards non-proprietary components in mobile/cellular network architectures. This includes the usage of \gls{cots} \glspl{sdr}. To this end, \citeauthor{upadhyaya2022prototyping}~\cite{upadhyaya2022prototyping} describe how to build an \gls{sdr}-based testbed supporting the \gls{oran} architecture and interfaces. This is accomplished by introducing standard-compliant open-source software-defined \glspl{ran}, such as srsRAN and \gls{oai}~\cite{nikaein2014openairinterface}. In~\cite{CFRAN}, a \gls{cf} implementation of \gls{oran} is presented, coined CF-RAN. They identify current implementation challenges such as time-frequency synchronization and reciprocity calibration, as handled in \cref{sec:calibrate_sync} in this work.  
Next to open interfaces, open commercial-grade software needs to be developed. In~\cite{agora}, a real-time \gls{mmimo} \gls{bb} processing software is presented, which is published in open-source. M-Cube~\cite{zhao2020m} is an open hardware and software for a \gls{mmwave} \gls{mmimo} \gls{sdr}, where the board schematics, real-time phased array controller code, interfaces, and examples are all available.

\subsubsection{Open Data}
Accompanying open testbed infrastructure and interfaces, open tools to process the generated (raw) data and the data itself should be accessible. In addition, the data ought to satisfy the FAIR principles~\cite{wilkinson2016fair}, i.e.,  findability, accessibility, interoperability, and reusability. This entails that the data is accompanied by well-described metadata, documentation, and (preferably) open licenses. This information, together with the data, can then be hosted on an open data platform, \update{such as} IEEE~DataPort, Open~Data~on~AWS, Kaggle and data.europa.eu (DEU).

\textit{Examples in the field.}
RenewLab~\cite{renewlab} is an open toolbox for processing \gls{mmimo} signals, used in the POWDER testbed~\cite{POWDER}. DeepSense~\gls{6g} \cite{alkhateeb2022deepsense} and LuViRA~\cite{LuViRA} are works describing measurements, including multi-modal sensors and well-described measurement campaigns, accompanied by an open dataset. All data sets generated by the COSMOS~\cite{raychaudhuri2020challenge} testbed are published on their website. Due to the increase in available \gls{6g} simulators, datasets including ray-tracing-based results are being published, e.g., DeepMIMO~\cite{DBLP:journals/corr/abs-1902-06435} and the ViWi dataset~\cite{Alrabeiah19}.

\section{Testbed Approaches By Example}\label{sec:testbeds} %

In this section, we address the aforementioned challenges and offer four examples of testbeds designed to meet specific testing and verification needs in emerging \gls{6g} systems. It is important to note that each of these presented testbeds \update{addresses different requirements}, leading to different implementations \update{and design choices}. To gain a better understanding of the discussed testbeds and their capabilities, see~\cref{tab:overview-testbeds}. For a comprehensive comparison of additional \gls{6g} testbeds, you can find relevant information in~\cite{OnTheRoadTo6G}.

\begin{table*}[h]
    \centering
    \caption{Summary of the presented testbeds, describing the testbed level, number of total antenna elements, bandwidth, carrier frequency and \gls{6g} functionality. A comprehensive overview of other \gls{6g} testbeds can be found in~\cite[Table~XI and~XII]{OnTheRoadTo6G}.}\label{tab:overview-testbeds}
    \begin{tabular}{l  r r l l l l l l l l l}
    \toprule
          &  & & & \multicolumn{6}{c}{Capabilities}\\
         Ref. & Antennas & Bandwidth & Frequency& \rotatebox{90}{Communication} & \rotatebox{90}{\acrshort{wpt}} & \rotatebox{90}{Localisation} & \rotatebox{90}{Open-Source} & \rotatebox{90}{\acrshort{dmimo}} & \rotatebox{90}{Testbed Level}\\ \toprule
         LuMaMi~\cite{7931558}&  100 & \SI{20}{\mega\hertz} & \SI{1.2}{\giga\hertz}-\SI{6}{\giga\hertz}& \yes & \no & \yes & \no & \no & L4\\
         LuMaMi28~\cite{chung2021lumami28} &  16 & \SI{20}{\mega\hertz} & \SI{27.95}{\giga\hertz} & \yes & \no & \yes & \no & \no & L4\\
         LuLIS (under dev.) &  256-1024 & \SI{100}{\mega\hertz} & \SI{3.7}{\giga\hertz} & \yes & \yes & \yes & \yes & \yes & L4\\
         KULMaMi~\cite{8254817} & 64 & \SI{20}{\mega\hertz} & \SI{400}{\mega\hertz}-\SI{4.4}{\giga\hertz}& \yes & \no & \yes & \no & \no & L4 \\
         Techtile~\cite{techtile} (under dev.) & 280 & \SI{10}{\mega\hertz} & \SI{70}{MHz}-\SI{6}{GHz} & \yes & \yes & \yes & \yes & \no & L4\\
         Chalmers D-MIMO~\cite{Sezgin2018} & 12 & \SI{40}{\mega\hertz} & \SI{2.35}{GHz} & \yes & \no & \yes & \no & \yes & L3\\ %
         Chalmers MATE~\cite{Buisman2018a} & 16 & \SI{1}{\giga\hertz} & \SI{28}{GHz} & \yes & \no & \no & \no & \no & L1/L3\\ %

         \bottomrule
    \end{tabular}
\end{table*}

\subsection{Scalable Low-Cost \Gls{dmimo} Testbed} \label{sec:Techtile}

\begin{figure*}[!htb]
      \begin{subfigure}[t]{0.39\textwidth}
         \centering
         \includegraphics[height=1.5in]{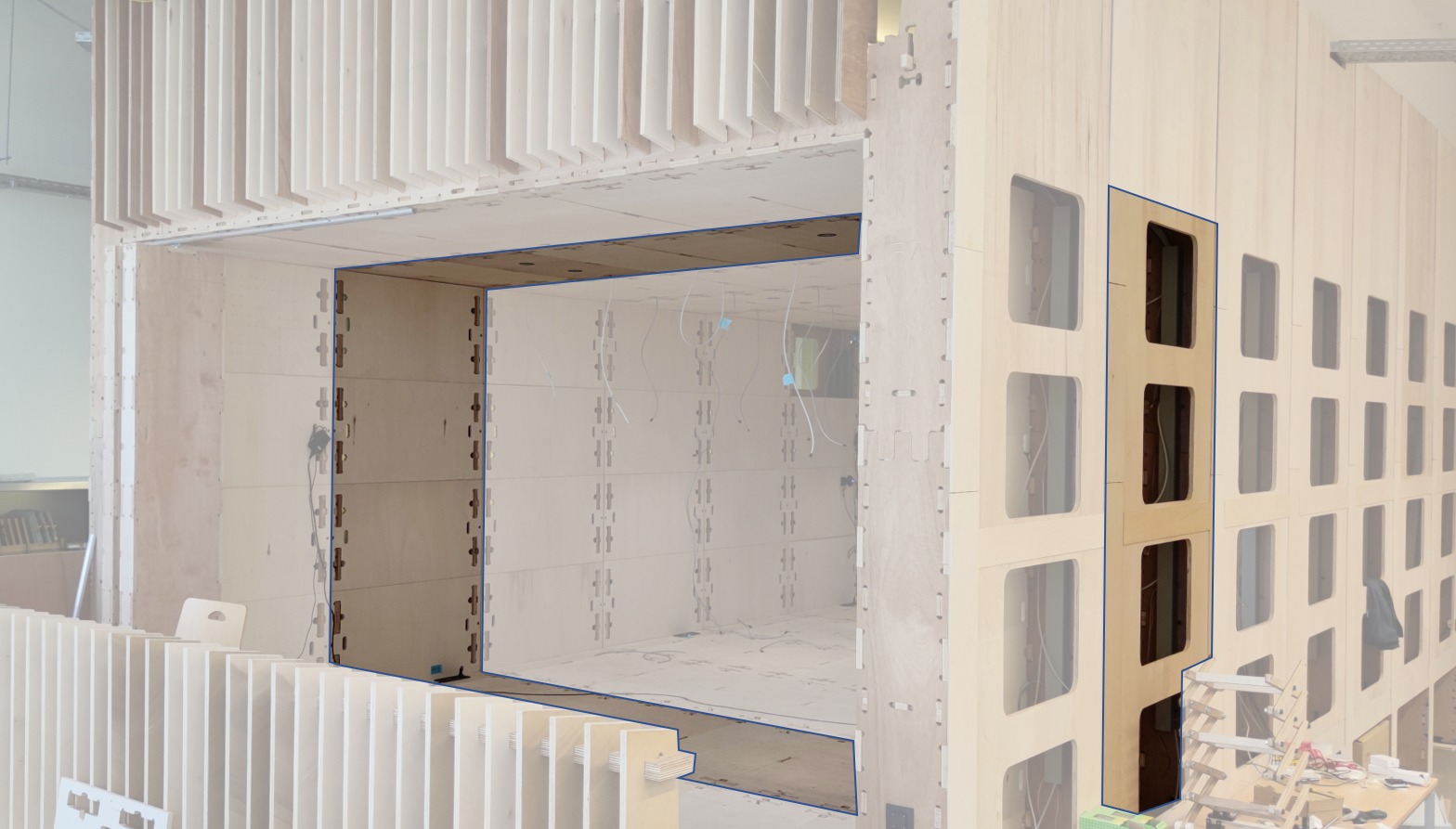}
         \caption{Dimensions: \SI{2.4}{m} (H), \SI{4}{m} (W), \SI{8}{m} (B).}
    \end{subfigure}\hfill%
    \begin{subfigure}[t]{0.19\textwidth}
    \centering
    \includegraphics[height=1.5in]{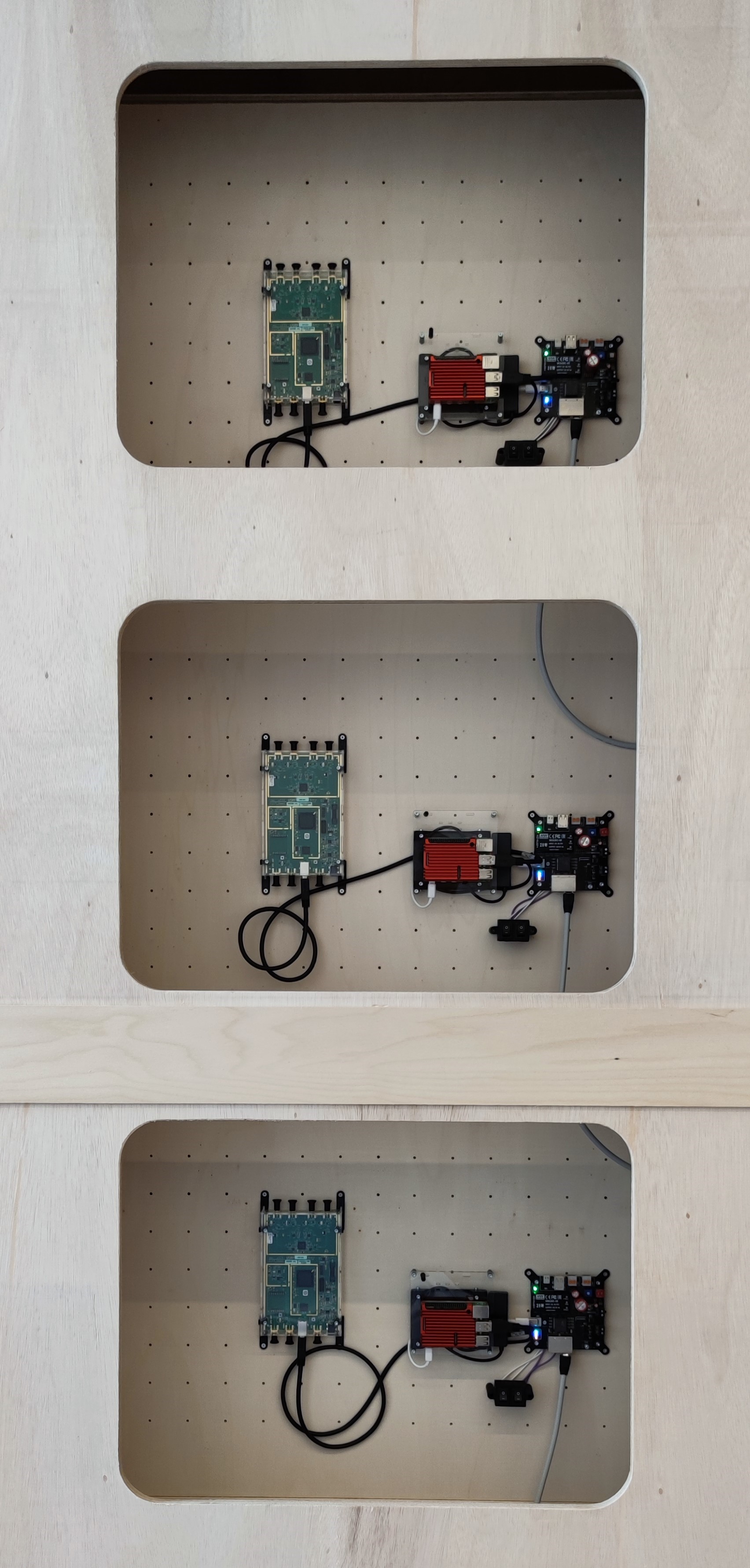}%
    \caption{The back of three tiles.}
    \end{subfigure}\hfill%
        \begin{subfigure}[t]{0.19\textwidth}
      \centering
      \includegraphics[height=1.5in]{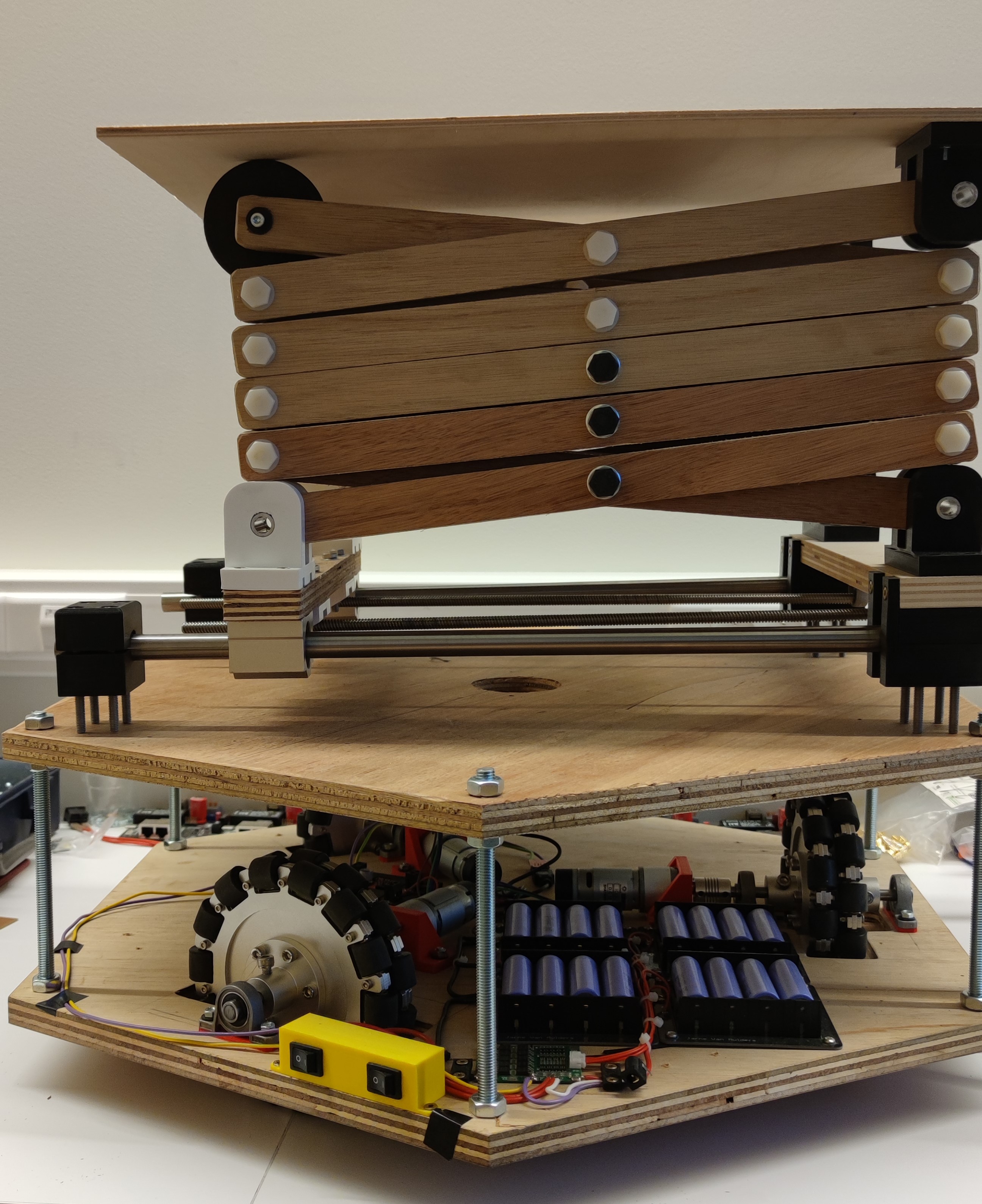}
      \caption{Techtile rover.}\label{fig:techtile-rover}
 \end{subfigure}\hfill%
    \begin{subfigure}[t]{0.19\textwidth}
       \centering
    \includegraphics[height=1.5in]{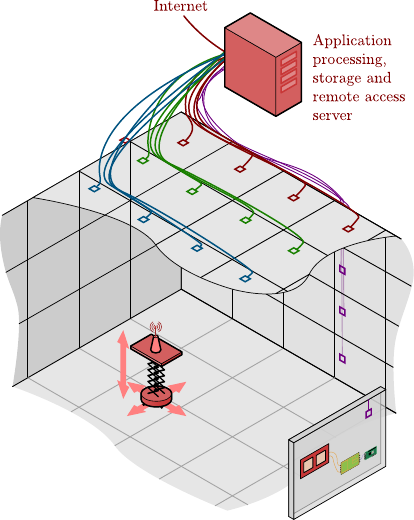}
    \caption{Schematic representation of the testbed.}
    \end{subfigure}
    \caption{The Techtile Testbed~\cite{techtile}, including the \gls{ue} rover (c). (b) shows the default setup for each tile, i.e., an \gls{sdr} (USRP B210), processing unit (Raspberry~Pi 4) and power supply with \gls{poe}. Each tile is connected to the central unit with an Ethernet cable, providing both power and data.}\label{fig:picture-techtile}
\end{figure*}

To study the impact on the architecture and performance of \gls{dmimo} with low-cost \gls{cots} components, Techtile was constructed~\cite{techtile}. The room-sized testbed consists of \num{140} tiles embedding computing devices, \glspl{sdr}, and sensors. The testbed and its architecture is illustrated in~\cref{fig:picture-techtile}. Techtile is fundamental in identifying and addressing key challenges in order to progress towards practically deployable \gls{dmimo}/\gls{cf} systems. These challenges stem from the high number of diverse resources hosted in the infrastructure and the novel application requirements.

\paragraph*{Evaluation of co-located vs. distributed APs}
As illustrated in~\cref{fig:picture-techtile}, all tiles are connected via Ethernet cables to a central server. By doing so, the physical topology allows emulating different topologies, e.g., daisy chain. This feature facilitates exploring different algorithms related to data transfer, as elaborated in Section~\ref{sec:co_vs_d_mimo}. 
By increasing the number of antenna elements (\num{280} elements), the users are expected to operate closer to the antenna near-field in \gls{dmimo} networks. With the physical large aperture of the testbed, this near-field operation can be investigated.

\paragraph*{Synchronization and Calibration}
Techtile currently supports Ethernet-based and dedicated cable-based synchronization. Fine-grained frequency, phase and time synchronization is obtained by distributing a \SI{10}{\mega\hertz} and \gls{pps} signal to all tiles via NI~Octoclocks. On top of this, the Ethernet infrastructure supports \gls{ptp} for sub-microsecond time synchronization.

\paragraph*{UE Data collection}
A rover with a scissor lift is used to automatically move \glspl{ue} in a 3D~space (\cref{fig:techtile-rover}). The rover uses multiple sensors to perform object detection and localization~\cite{techtile}.

\paragraph*{Open Testbed}
All developed software and hardware is available and documented on our GitHub webpage (\url{www.github.com/techtile-by-dramco}).

\subsection{Sigma-delta radio-over-fiber enabled D-MIMO testbed}\label{sec:DRoF_testbed}
As discussed in Section~\ref{sec:calibrate_sync}, one of the main challenges in realizing \gls{dmimo} systems is to maintain precise synchronization between remote \gls{csp}. \Gls{rof} is a technique that has been explored for realization of phase coherent synchronization in distributed antenna systems. Both digital- and analog implementations of \gls{rof} exist. Analog implementations, where \update{the RF signal amplitude modulates} the optical signal intensity, \update{have} the benefit of simple \gls{rf} circuit implementation but \update{suffer} from limited dynamic range and complex optical implementations \cite{breyne2017comparison}. Digital \gls{rof} implementations, on the other hand, are robust against optical imperfections and benefit from the rapid evolution of standardized low-cost and ultra-high-speed digital optical interconnect solutions for data centers.

\begin{figure*}
    \centering
    \begin{minipage}{.45\linewidth}
    \begin{subfigure}[t]{0.9\textwidth}
    \centering
        \includegraphics[width=\textwidth]{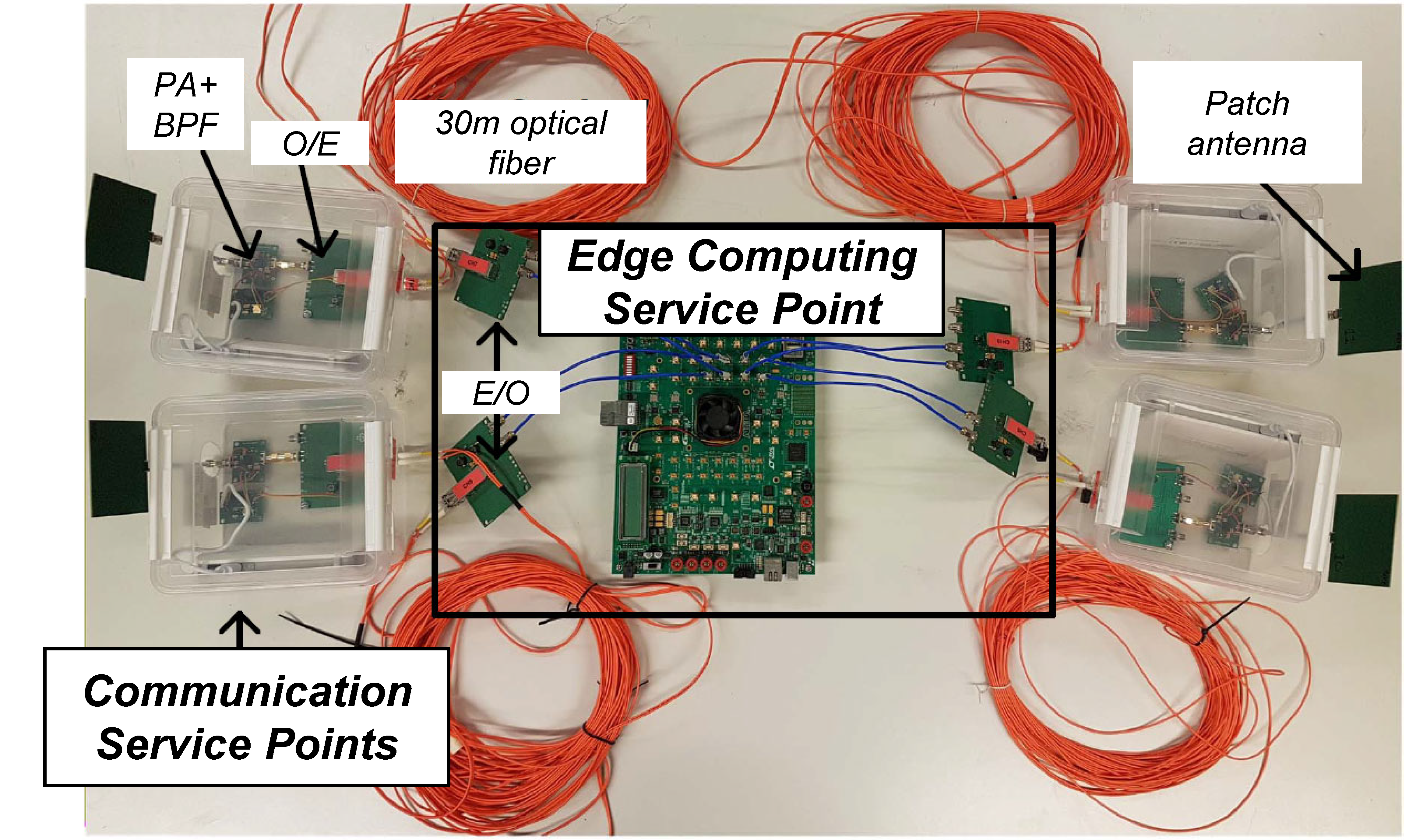}
        \caption{\Acrfull{ecsp}}
        \label{fig:SDoF_Testbed}
    \end{subfigure} \\
    \begin{subfigure}[b]{0.9\textwidth}
         \centering
         \includegraphics[width=\textwidth]{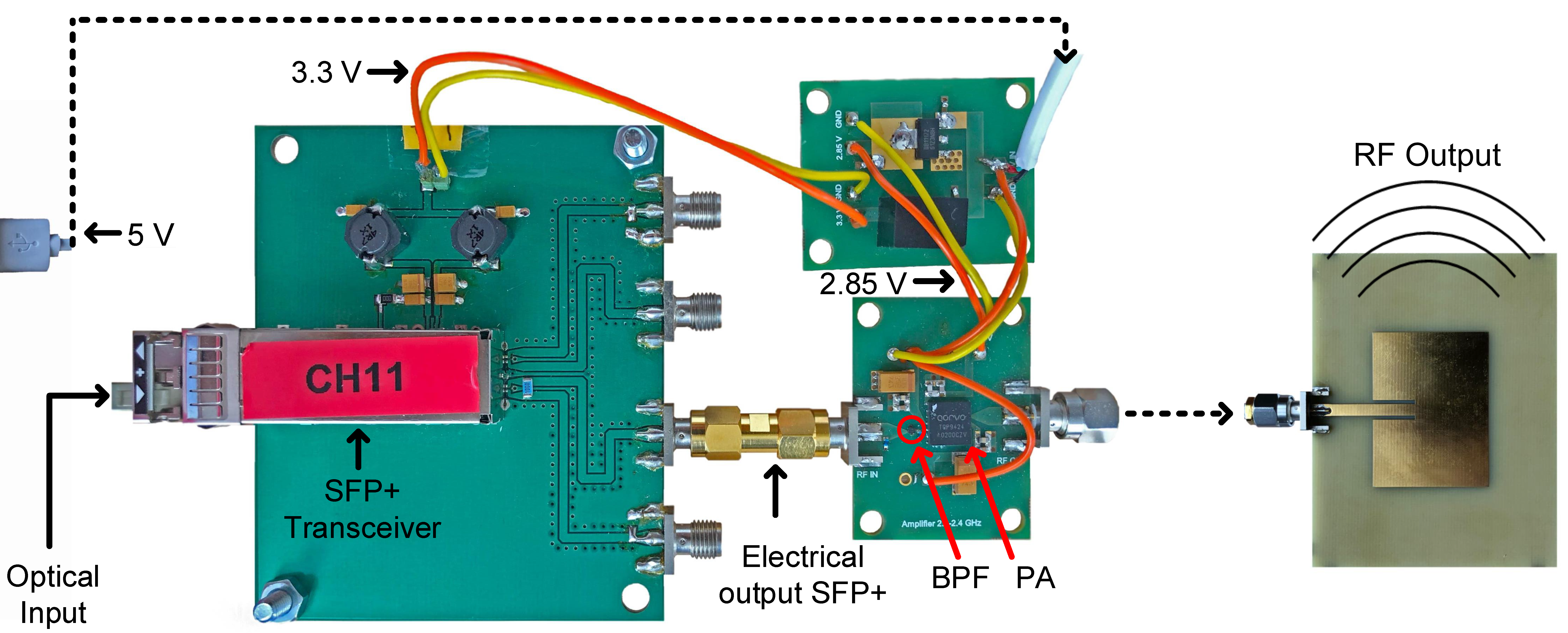}
        \caption{\Acrfull{csp}}
        \label{fig:SDoF_AP}
    \end{subfigure}
    \end{minipage}
    \begin{minipage}{.45\linewidth}
    \begin{subfigure}[b]{0.9\textwidth}
         \centering
         \includegraphics[width=0.85\textwidth]{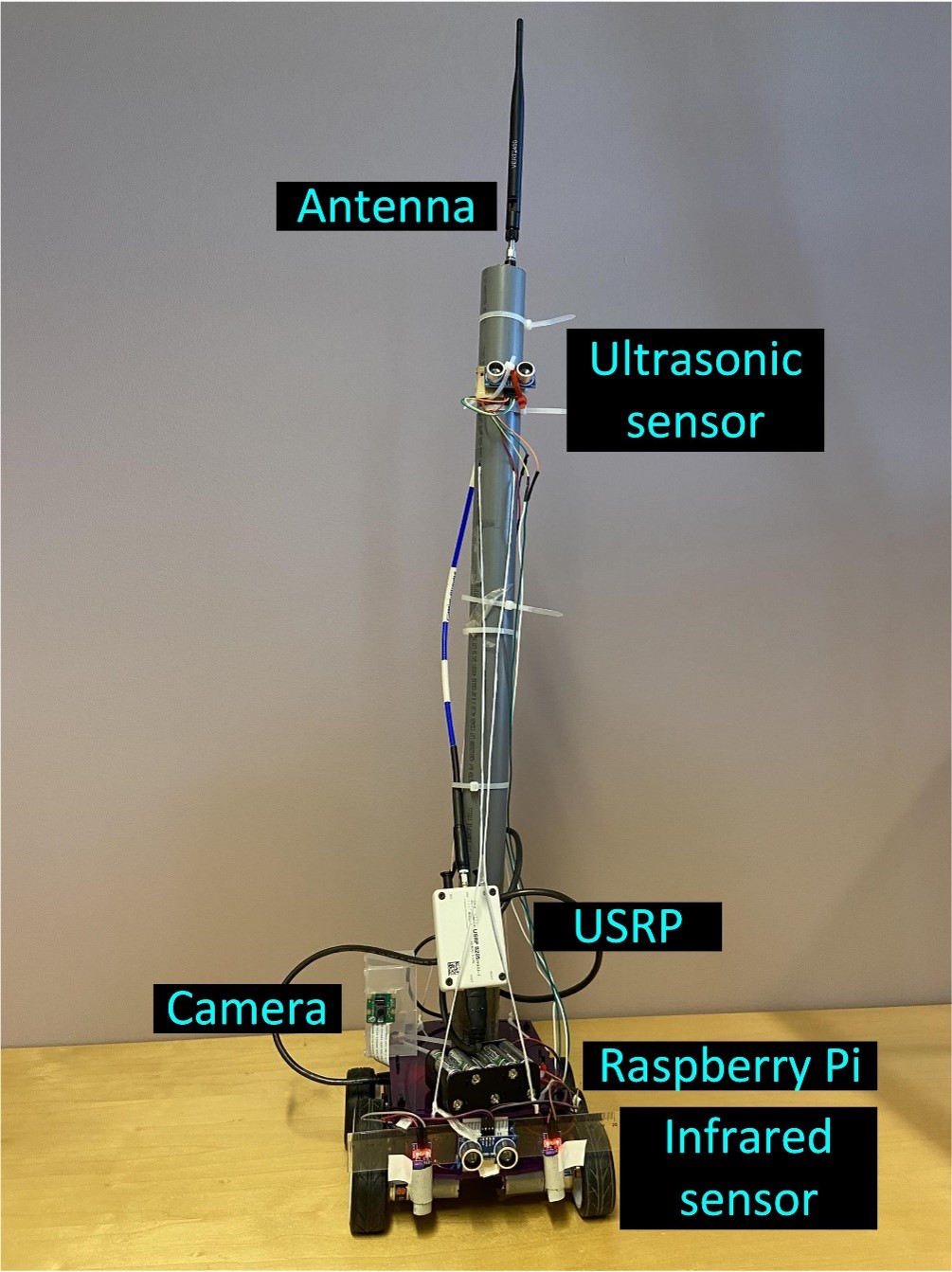}
        \caption{\Gls{ue} robot}
        \label{fig:SDoF_robot}
    \end{subfigure}
    \end{minipage}
    \caption{Photo illustrating the key components of the sigma-delta radio-over-fiber testbed~\cite{Sezgin2018}. The testbed offers phase coherence across large distances, making it particularly suited for distributed MIMO communication and localization experiments.}
\end{figure*}

The \gls{sdof} technique is employed in~\cite{Sezgin2018} to realize a testbed suitable for \gls{dmimo} \ac{dl} experiments. As illustrated in~\cref{fig:SDoF_Testbed}, a \gls{ecsp} (in this case an \gls{fpga}) \update{feeds} up to \num{12}~remote \glspl{csp} through \SI{30}{\meter} \SI{10}{Gbit/s} \gls{sfp} optical links with individually coded bandpass \sigdel-coded \ac{rf} bitstreams. At the remote \gls{csp}, the \SI{2.35}{\giga\hertz} signal is recovered from the bitstream through a \gls{baw} bandpass filter, then amplified and fed to a patch antenna. As demonstrated in~\cite{Sezgin2018}, the testbed can support up to \SI{40}{\mega\hertz} bandwidth and maintains an \gls{rms} phase coherence below \SI{2.1}{\degree} between the \glspl{csp}. 

\paragraph*{UE Data collection} 
To facilitate efficient collection of communication and channel data in realistic environments, the testbed has been extended with an \gls{ue} robot in~\cite{Bao21}, see~\cref{fig:SDoF_robot}. The robot is controlled by a Raspberry Pi, an Ettus USRP radio receiver and optical sensors. The robot is programmed to follow a line on the floor/ground and, at intervals marked by perpendicular lines, collects over-the-air data and transmits it to the \gls{ecsp} using Wi-Fi for communication signal processing. 

\begin{figure*}
    \centering
    \begin{subfigure}[t]{0.39\textwidth}
    \centering
        \includegraphics[height=4.5cm]{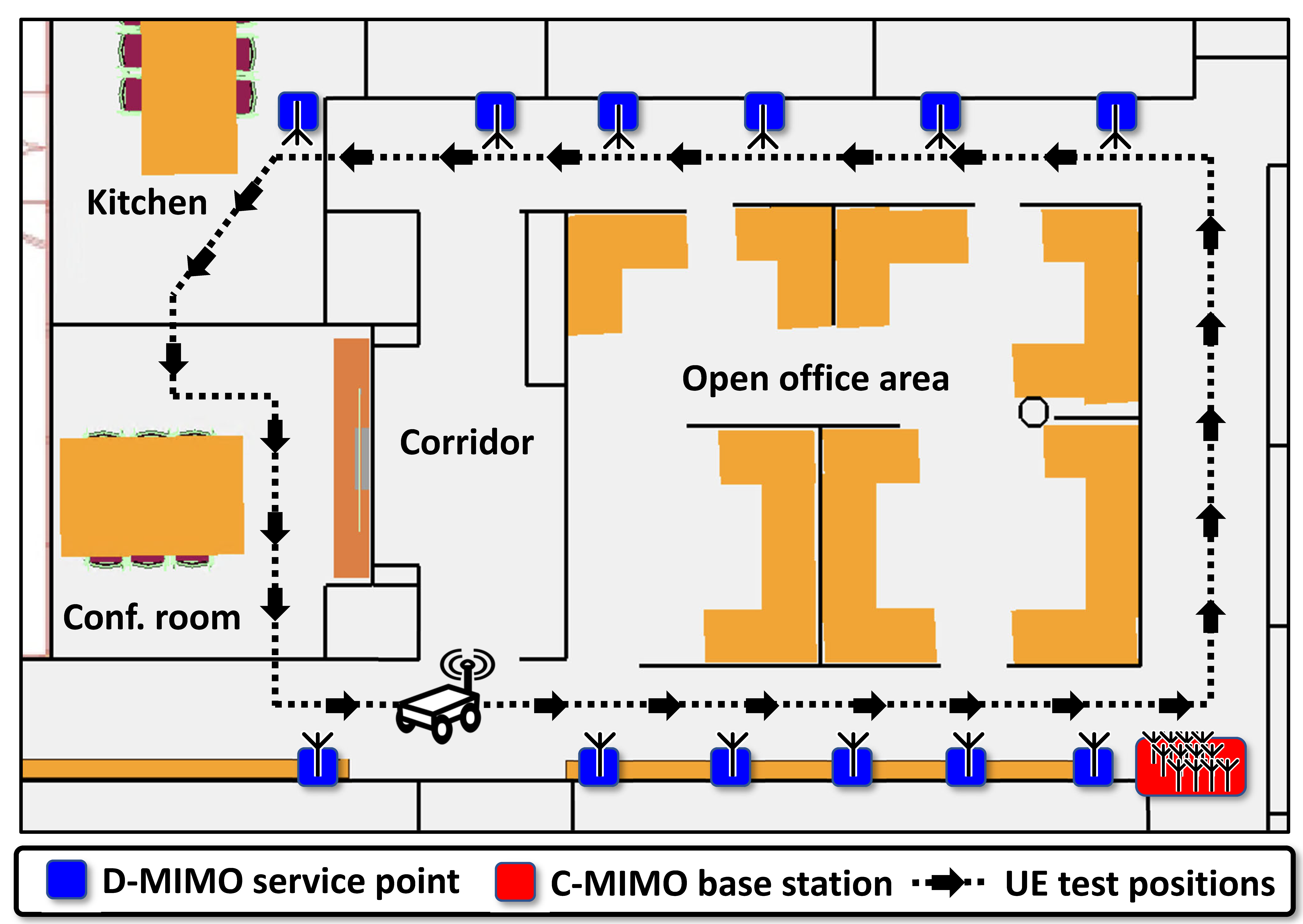}
    \caption{Measurement scenario.}
    \label{fig:SDoF_IndoorOffice}
    \end{subfigure}\hfill%
    \begin{subfigure}[t]{0.59\textwidth}
    \centering
    \centering
    \includegraphics[height=4.5cm]{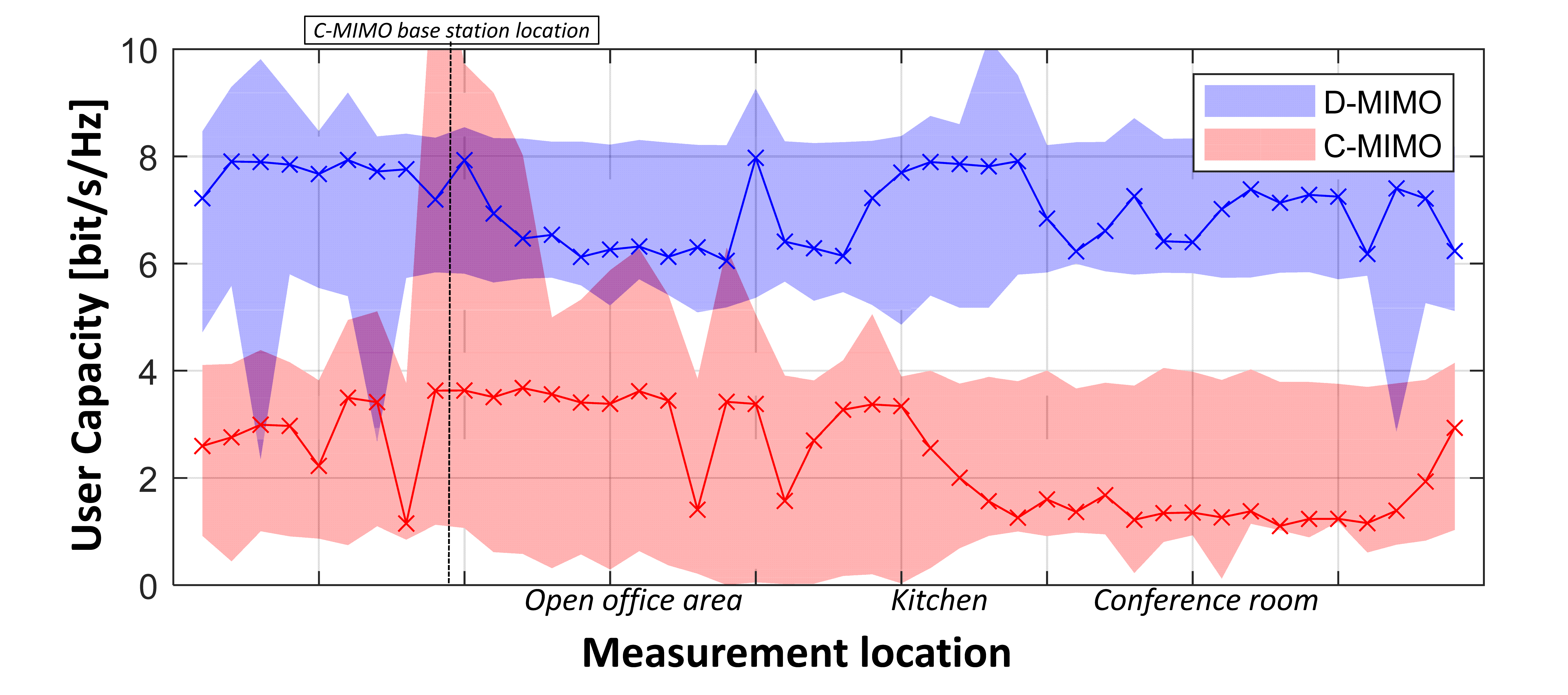}
    \caption{Estimated median user capacity variations in a four-user MIMO scenario. The colored regions represent the $25^{\mathrm{th}}/75^{\mathrm{th}}$ percentile.}
    \label{fig:SDoF_IndoorOffice_Capacity}
    \end{subfigure}
    \caption{Evaluation of distributed vs. co-located MIMO communication capacity for a in an indoor office environment using an automated robot for UE data collection \cite{Bao21}.}
\end{figure*}

\paragraph*{Evaluation of co-located vs. distributed \glspl{csp}}
In~\cite{Fager22,Bao21}, the testbed has been \update{used} to evaluate communication performance in a realistic indoor office environment, see~\cref{fig:SDoF_IndoorOffice}. 
The optical interconnects offer great flexibility in the placement of the \glspl{csp}, which makes the testbed particularly suitable for experimental evaluation of distributed versus co-located MIMO as discussed in~\cref{sec:co_vs_d_mimo}. In one of the testbed experiments, over-the-air channel information was collected and used to evaluate the multi-user MIMO capacity for a randomized location of four simultaneous users. As demonstrated from the results in~\cref{fig:SDoF_IndoorOffice_Capacity}, the \gls{dmimo} (i.e., \gls{6g}) configuration offers much more uniform user capacity, compared to the co-located MIMO (i.e., 5G) case~\cite{Bao21}. These results confirm the theoretical predictions in, e.g.,~\cite{interdonato2019ubiquitous}. 

\paragraph*{Localization}
The accurate phase coherence between different \glspl{csp} enables precise localization to be performed. In~\cite{keskin2022localization}, the testbed was used to demonstrate a localization accuracy of below~\SI{0.2}{\meter} over an area of~\SI{100}{\square\meter}, in agreement with the theoretical bounds at most measurement locations. These results also validate the feasibility of \gls{dmimo} for joint communication and localization applications.
\subsection{Millimeter-wave MIMO testbed (MATE)}
\label{sec:MATE_testbed}

\begin{figure}
    \centering
    \includegraphics[width=0.8\linewidth]{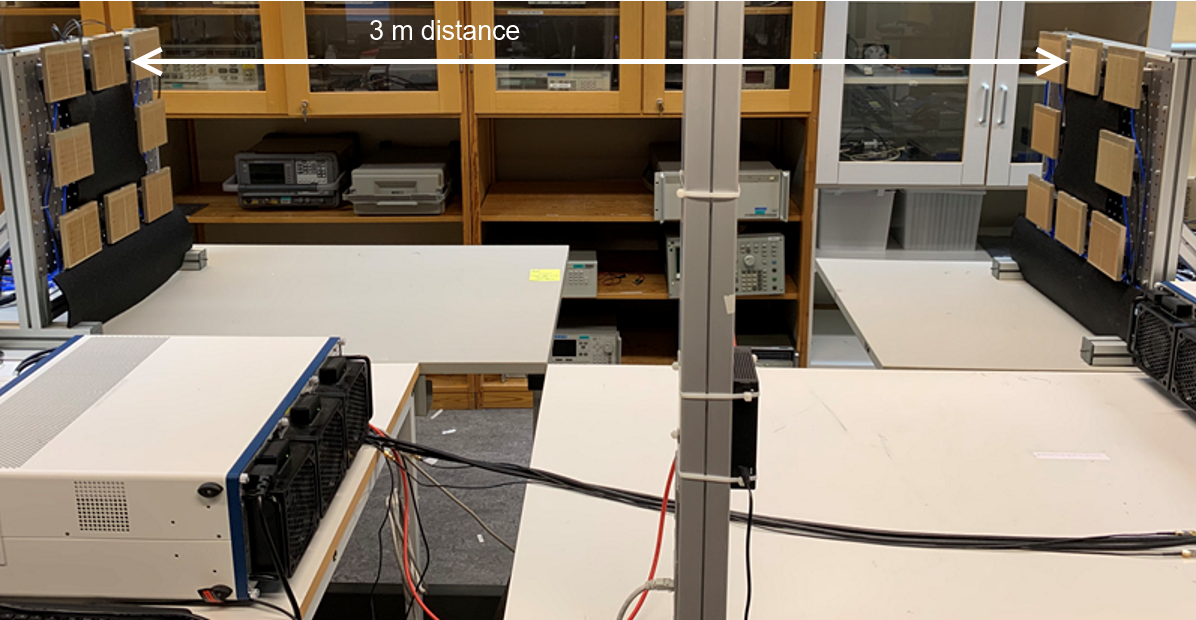}
    \caption{The \gls{mate} testbed~\cite{Buisman2018a}, in a configuration with 8~\gls{tx} and 8~\gls{rx} highly directive antennas. Various kinds of research, on hardware, antennas, imperfection mitigation, and \gls{mimo} communication or radar processing, is enabled through the testbed's online web interface.}
    \label{fig:Mate}
\end{figure}

When designing the \emph{MATE} \gls{mmwave} testbed, it was intended to \update{be fabricated with} \gls{cots} components, with the \update{flexibility} to replace those with our own designs. A main focus was research on calibration and the mitigation of hardware imperfections, in particular \gls{ota}. An intended application was MIMO communication, mainly MIMO links, but potentially also multiuser MIMO, for \glspl{mmwave}. Radar signal processing was not a main goal, but a potential study topic. Another requirement was that it should be easily accessible \update{for people that have not spent a lot of time in the lab environment}. 
These requirements translated into a testbed that operates \update{over the range} \SIrange{28}{31}{GHz}, with \SI{1}{GHz} analog bandwidth per transmitter or receiver. \Gls{mate} supports up to 18~channels, which can be used in various configurations, with up to 16~transmitters and up to 9~receivers. The \gls{mate} baseband transmitter is implemented with \glspl{dac}, \glspl{fpga}, sample clocks and triggering. To ensure the baseband signals are coherent, one sample clock is distributed to all \glspl{dac}. The baseband receiver is similarly implemented, independent of the transmitter, with a single clock distributed to all \glspl{adc}, The baseband hardware (\cref{fig:Mate}) consists of National Instruments PXIe chassis (NI PXIe-1085), reference clock/trigger modules (NI PXIe-6674T), \glspl{dac} (Active Technologies AT-1212), \glspl{adc} (NI-5771) and \glspl{fpga} (PXIe-7975R). One or two embedded controllers (NI PXIe- 8880) control the system.

The testbed is controlled by a web interface, where a MATLAB client can be used to easily upload baseband data vectors to all antennas. These are up-converted to \SI{28}{GHz}, transmitted over the TX antenna array, and received by the RX array. The receiver \update{down-converts} to baseband, and \update{sends} the data vectors back to MATLAB, over the web. 
The entire process takes about 8 seconds.  
Over time, many types of experiments have been performed using \gls{mate}. The large bandwidth enables research of a nature not available in most testbeds. To give a few examples, low-profile high-gain antennas were designed and manufactured~\cite{Shi2019} (\cref{fig:Mate}). 
Calibration was studied in~\cite{Buisman2018a}, and \gls{ota} calibration in~\cite{Buisman2022}. Mitigation and modeling of hardware imperfections \update{were} studied in~\cite{Aghdam2018}, and it was \update{adapted} to nonlinear distortion in~\cite{Nachouane2021}. Multi-user interference, and the effect on the signal quality, was studied in~\cite{Buisman2018}. %

\subsection{Real-time \ac{mmimo} testbed: from centralized to distributed architecture}

\begin{figure*}[tb!]
  \centering
  \begin{subfigure}[b]{0.5\textwidth}
    \centering
    \includegraphics[height=4cm]{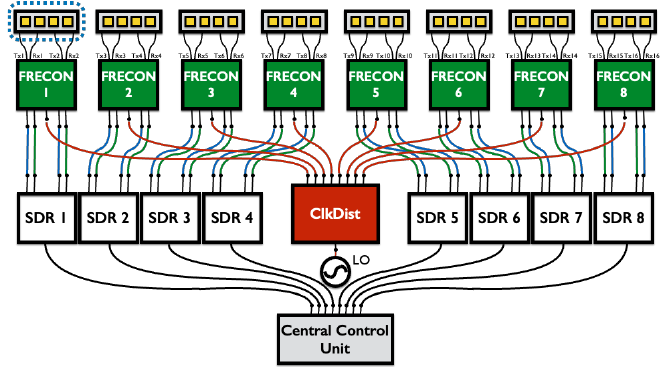}
  \end{subfigure}\hfill%
  \begin{subfigure}[b]{0.5\textwidth}
    \centering
    \includegraphics[height=4cm]{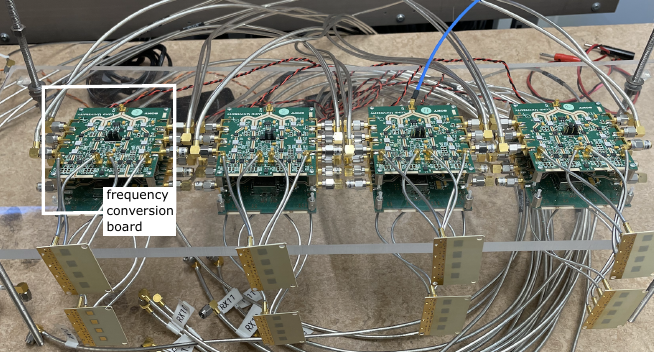}
  \end{subfigure}
  \caption{LuMaMi with 8 frequency conversion boards to operate 16-antenna fully-digital beam-forming at \SI{27.95}{GHz}~\cite{chung2021lumami28}. LuMaMi generates \SI{2.45}{GHz} \gls{if} signals which are up-converted by image-rejection mixers to the operating frequency. The \SI{25.5}{GHz} \gls{lo} signal is amplified and distributed to the \num{16}~mixers through a dedicated clock distribution board.}%
  \label{fig:lumami28}
\end{figure*}

\begin{figure}[tb!]
  \centering
  \includegraphics[width=0.8\linewidth]{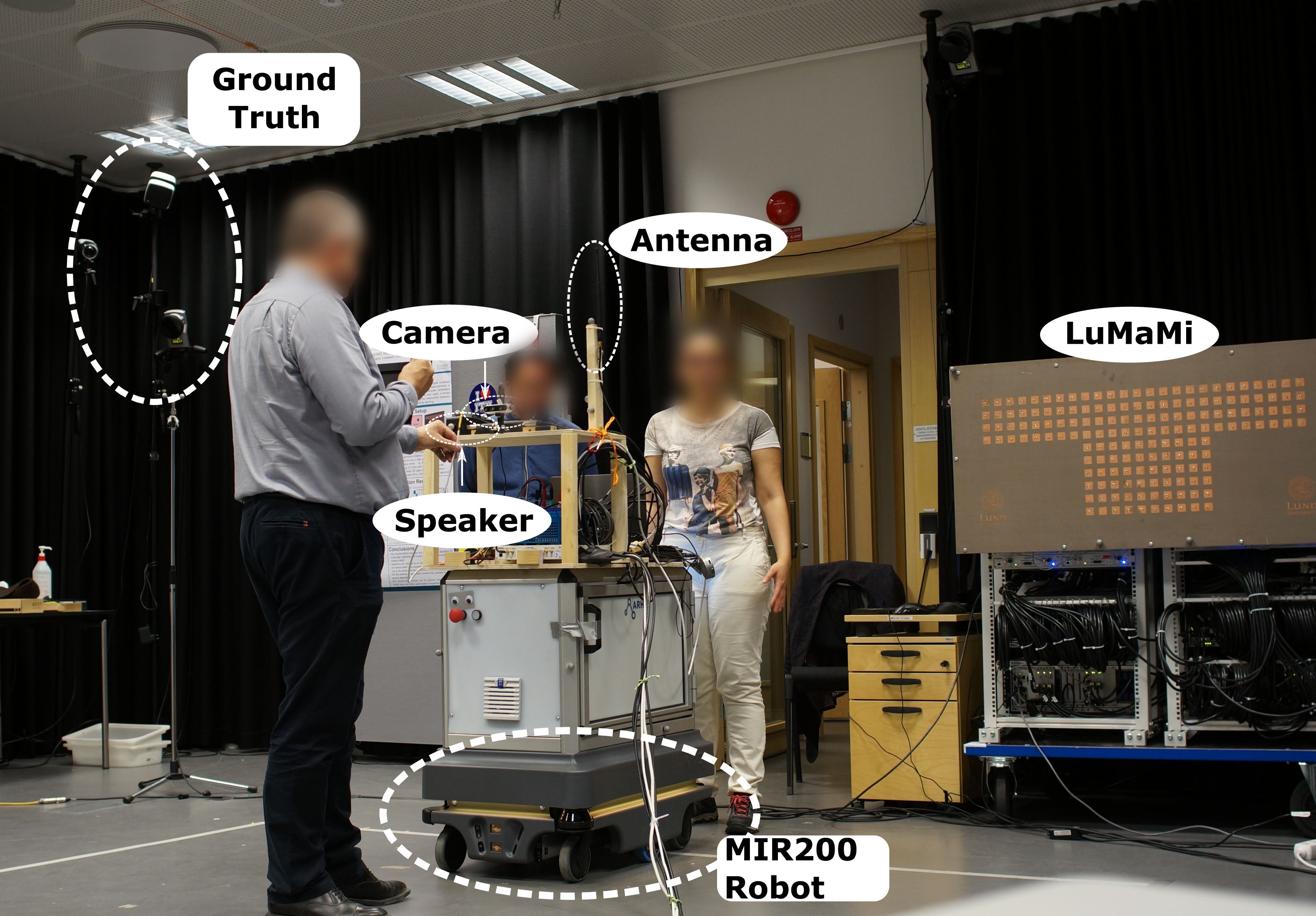}
  \caption{LuMaMi used in measurement campaign to collect real-life data for high-accuracy and multi-sensory positioning~\cite{LuViRA}. Mobile MIR200 robot carrying sensors/transmitters for audio, image, and radio propagation data collection. The position ground truth is provided by a high-precision motion capture system with \num{18}~networked high-speed infrared cameras.}
  \label{fig:luvira}
\end{figure}

\begin{figure*}
    \centering
    \begin{subfigure}[b]{0.4\textwidth}
      \centering
      \includegraphics[height=6cm]{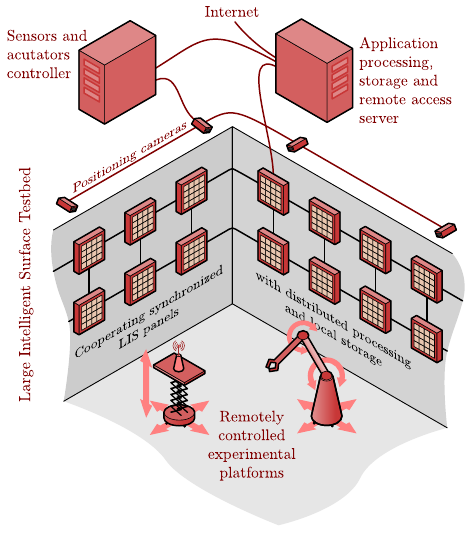}
    \caption{}
    \label{fig:LuLIS}
    \end{subfigure}\qquad%
\begin{subfigure}[b]{0.4\textwidth}
  \centering
  \includegraphics[height=5cm]{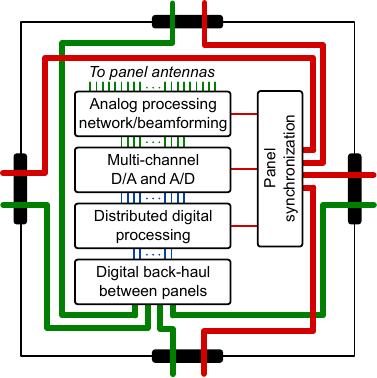}
    \caption{}
    \label{fig:LuLIS_details}
\end{subfigure}
    \caption{(a) Software defined \gls{lis} testbed under development at Lund University (LuLIS). Cooperating synchronized panels, based on AMD/Xilinx RFSoC technology. (b) Overview of the internal architecture of the \gls{lis} testbed panels.}
\end{figure*}

The LuMaMi testbed~\cite{7931558} at Lund University is a real-time OFDM-based \gls{mmimo} testbed, with 50~\glspl{sdr} from NI, each with two TX and RX chains and Kintex-7 \glspl{fpga}, transmitting and receiving coherently and simultaneously on 100 antenna elements. The frequency range of the \glspl{sdr} is \SIrange{1.2}{6}{GHz}, and the current setup is tailored for \SI{3.7}{GHz}, with the centralized antenna array tuned to that frequency. With the \glspl{fpga} in the \glspl{sdr}, together with four additional \glspl{fpga} for centralized processing, LuMaMi is capable of real-time operation of a 30.72~MS/s LTE-like air interface in TDD mode, with 1200 active subcarriers across \update{a} 20~MHz bandwidth. Due to the TDD-operation and reliance on reciprocity for channel knowledge and beamforming, the testbed also contains the necessary functionality for reciprocity calibration~\cite{Vieira2017reciprocity}. 
LuMaMi has been used in several different roles, testing \update{the} efficiency of spatial multiplexing~\cite{SErecord2016}, testing real-time processing in high mobility environments~\cite{Harris2017a}, and exploring \gls{mmimo} technology for new services in the \SI{28}{GHz} band, as illustrated in~\cref{fig:lumami28}~\cite{chung2021lumami28}. \update{Because of} the \gls{sdr} flexibility, \update{the testbed} has also been used as a robust low-latency wireless link in cloud control of robots~\cite{skarin2018towards}, a channel sounder in industrial environments~\cite{5GSMARTD4.2}, and \update{for} collecting audio/video/radio propagation data-sets for machine-learning based positioning services (illustrated in~\cref{fig:luvira})~\cite{LuViRA}. 
Lund University is currently building on the experience with LuMaMi and \update{is} developing a distributed and modular testbed, where panels of 16-antenna arrays with localized processing capabilities are distributed in the environment, as illustrated in~\cref{fig:LuLIS}, and together operate coherently with a bandwidth of \SI{100}{MHz} in the \SI{3.7}{GHz} band. An overview of the internal architecture of the panels is shown in~\cref{fig:LuLIS_details}. The goal of this design is to further explore what can be achieved by exploiting spatial domain even further than what is done in traditional \gls{mmimo}. Based on RFSoC technology, the testbed is software controlled and supports real-time distributed processing, making it capable of both exploring many different service types, such as communication and positioning and also \update{performing} propagation measurements for the purpose of channel modeling, sensing and creation of datasets for machine-learning experiments.

\section{Future directions}
We identify two main future directions for \gls{6g} (and beyond) testbeds, and discuss these shortly below: the reduction of the complexity in the implementation of the first generation of \gls{6g} testbeds on the one hand, and extending their features and enabling the validation of new paradigms on the other hand.  

\subsection{Complexity reductions}
The development of the first generation of \gls{6g} testbeds most importantly targets the \textit{functional} possibility \update{for validating} new features and architectures of \gls{6g} technologies. In future upgrades, it should be a priority to achieve this with a \textit{low complexity implementation}. 
By shifting away from dedicated systems and adopting low-cost \gls{cots} products, these systems can be established and deployed in more labs, becoming more accessible. This accessibility can act as a catalyst for advancing innovative ideas in \gls{6g}, ultimately leading to valuable technologies. Additionally, since these systems must accommodate large \glspl{csp}, the technologies, and hardware used must be scalable in terms of cost, hardware/software complexity, and energy consumption. 
Also, challenges encountered in the actual deployment of new wireless technologies often only fully reveal themselves in the creation of real-life testbeds.
Moreover, solutions found and lessons learned from complexity and energy reductions in \gls{6g} testbeds may 'spill over' to actual \gls{6g} deployments.  
The design of efficient implementations of testbeds may hence be an important fuel of innovation by itself.  

\subsection{Support for new features and paradigms}
\subsubsection*{Superconvergence and co-created networks}
The vision for \gls{6g} \update{outlines} a broad trend of superconvergence of and in networks. The evolution towards \gls{multi-rat} support, in particular in combining solid connectivity in \update{bqnds} below \SI{20}{GHz} with sub-THz wide bandwidth channels, is potentially in the foreseeable future the only answer to the question of offering at the same time a very high throughput and high reliability. A 'tandem' operation, researched in the 6GTandem project~(\url{www.horizon-6gtandem.eu}), bears the potential to offer the best of two worlds. 
The validation of network-of-networks, operating both in licensed and unlicensed bands, and comprising terrestrial and non-terrestrial components, comes with enormous complexity. This will require a joint approach involving both network simulators~\cite{9363049} and real-life testbeds. The new paradigm of networks that are co-created by citizens and to a large extent may result in bottom-up deployments, still opens a whole new dimension of freedom and potential chaos. The validation and optimization of these approaches also raise the need for new features to be supported in testbeds.

\subsubsection*{Millimeter-wave and sub-THz extensions}
Another important future direction will be to extend current \gls{dmimo} testbeds to \gls{mmwave} and potentially sub-THz bands. \citeauthor{Moerman2022}~review the potential of \gls{mmwave} \gls{dmimo} and present an interesting testbed approach based on analog~\ac{rof}~\cite{Moerman2022}. In another recent work~\cite{Bao2023}, a flexible \gls{sdof} link architecture, suitable for \gls{mmwave} \gls{dmimo} testbed applications, is presented. 

\section{Conclusions}\glsresetall
As diverse as the applications and technologies envisioned for \gls{6g} networks, \update{so too} are the testbeds required to validate these technologies. This paper has introduced both different levels and various scopes for \gls{6g} testbeds. \update{The article} further focused on the main new capacity technology being \gls{dmimo}. It has discussed approaches to address scaling and distributing the number of antennas and processing over the network, under more realistic conditions and in terms of hardware \update{for studying} real-life propagation situations and real-time behavior. Critical deployment challenges, such as the need to achieve synchronization in distributed architectures, are encountered in the development of these testbeds. These require solutions that may also bring great value in actual \gls{6g} deployments. Hence, this work highlights and details how \gls{dmimo} \gls{6g} systems could be designed, implemented and deployed. 
The R\&D towards \gls{6g} is far from finished, and so is the work to develop adequate testbeds. We have outlined directions to reduce complexity and enrich features in future developments.

\section*{Acknowledgments}
\begin{itemize*}
    \item[] This project has received funding from the European Union’s Horizon 2020 research and innovation programme under the Marie Skłodowska-Curie grant agreement No 860023.
    \item[] Part of this research has been carried out in Gigahertz-ChaseOn Bridge Center in a project financed by Chalmers, Ericsson, Gotmic, Infineon, Kongsberg, Saab, and UniqueSec.
    \item[] This work was partially funded by the Swedish Research Council (grant VR-2019-05174).
    \item[] This work was partially funded by the Swedish Foundation for Strategic Research (Project ID CHl19-0001).
    \item[] 6GTandem has received funding from the Smart Networks and Services Joint Undertaking (SNS JU) under the European Union’s Horizon Europe research and innovation programme under Grant Agreement No~101096302. 
    \item[] The REINDEER project has received funding from the European Union’s Horizon 2020 research and innovation programme under grant agreement No.~101013425.
    \item[] BEYOND5 -- Building the fully European supplY chain on RFSOI, enabling New Domains for Sensing, Communication, 5G and beyond. H2020, Project No~876124. 
\end{itemize*}

\printbibliography

\end{document}